\documentclass[twocolumn,iop,apj]{aastex63}
\usepackage{amsfonts,amsmath,graphicx,natbib,url,hyperref,nicefrac}
\usepackage{amssymb, verbatim, subfigure, paralist, soul, comment}

\hypersetup{colorlinks=true, urlcolor=blue, citecolor=blue}

\usepackage{color}
\newcommand{\sprout}{\texttt{Sprout}}
\newcommand{\sm}[1]{\textcolor{black}{#1}}

\shorttitle{anisotropies in snr observations} 
\shortauthors{Mandal et al.}

\begin{document}

\title{Measurement of anisotropies in Supernova Remnant observations and their interpretation using numerical models}

\author[0000-0001-9484-1262]{Soham Mandal}
\affiliation{Department of Physics and Astronomy, Purdue University, 525 Northwestern Avenue, West Lafayette, IN 47907, USA}

\author[0000-0001-7626-9629]{Paul C. Duffell}
\affiliation{Department of Physics and Astronomy, Purdue University, 525 Northwestern Avenue, West Lafayette, IN 47907, USA}

\author[0000-0002-1633-6495]{Abigail Polin}
\affiliation{The Observatories of the Carnegie Institution for Science, 813 Santa Barbara St., Pasadena, CA 91101, USA}
\affiliation{TAPIR, Walter Burke Institute for Theoretical Physics, 350-17, Caltech, Pasadena, CA 91125, USA}
\affiliation{Department of Physics and Astronomy, Purdue University, 525 Northwestern Avenue, West Lafayette, IN 47907, USA}

\author[0000-0002-0763-3885]{Dan Milisavljevic}
\affiliation{Department of Physics and Astronomy, Purdue University, 525 Northwestern Avenue, West Lafayette, IN 47907, USA}
\affiliation{Integrative Data Science Initiative, Purdue University, West Lafayette, IN 47907, USA}

\email{mandal0@purdue.edu}

\begin{abstract}

Supernova remnants (SNRs) exhibit varying degrees of anisotropy, which have been extensively modeled using numerical methods. We implement a technique to measure anisotropies in SNRs by calculating power spectra from their high-resolution images. To test this technique, we develop 3D hydrodynamical models of supernova remnants and generate synthetic x-ray images from them. Power spectra extracted from both the 3D models and the synthetic images exhibit the same dominant angular scale, which separates large scale features from small scale features due to hydrodynamic instabilities. The angular power spectrum at small length scales during relatively early times is too steep to be consistent with Kolmogorov turbulence, but it transitions to Kolmogorov turbulence at late times.  As an example of how this technique can be applied to observations, we extract a power spectrum from a \textit{Chandra} observation of Tycho's SNR and compare with our models. Our predicted power spectrum picks out the angular scale of Tycho's fleece-like structures and also agrees with the small-scale power seen in Tycho. We use this to extract an estimate for the density of the circumstellar gas ($n \sim 0.28/\mathrm{cm^3}$), consistent with previous measurements of this density by other means. The power spectrum also provides an estimate of the density profile of the outermost ejecta. Moreover, we observe additional power at large scales which may provide important clues about the explosion mechanism itself.

\end{abstract}


\section{Introduction}  \label{sec:intro}

Observations of Supernova Remnants (SNRs) reveal that they exhibit varying degrees of departure from spherical symmetry. These anisotropies range from triaxial near-ellipsoid shapes \citep[SN 1987A;][]{Kjaer+2010A&A,McCray+2016ARA&A}, barrel shaped structures \citep[W49B;][]{Lopez+2013ApJ}, and large scale asymmetric protrusions \citep[Cassiopeia A;][]{Fesen+2016ApJ} to small-scale anisotropies including `fleece'-like structures distributed over an overall spherical shape, e.g., Kepler's SNR \citep{Reynolds+2007ApJ}, Tycho's SNR \citep{Warren+2005ApJ} and SN 1006 \citep{Winkler+2003ApJ}. They are attributed to a multitude of phenomena, including Rayleigh-Taylor Instability \citep[RTI;][]{Chevalier+1978ApJ,Velazquez+1998A&A}, the presence of dense clumps in the circumstellar medium \citep[CSM;][]{Celli+2019MNRAS,Sano+2020ApJ_a,Sano+2020ApJ_b}, inherent anisotropies in the SN ejecta due to neutrino driven convection and radioactive heating by $^{56}$Ni clumps \citep{Wongwathanarat+2015A&A,Wongwathanarat+2017ApJ,Orlando+2016ApJ,Orlando+2021A&A,Gabler+2021MNRAS}, standing accretion shock instability \citep{Blondin2005,Blondin+2007Nature,Iwakami+2008ApJ}, jets \citep{GC+2014ApJ,Bear+2017MNRAS} etc. Thus, broadly speaking, the features observed in SNRs may be due to anisotropies endemic to the SN explosion itself (such as jets, Ni clumps, or presence of a companion), and/or due to interaction of the ejecta with the CSM, which may provide anisotropies externally. 

Previous studies on the nature of anisotropies in SNRs \citep{Ferrand+2019ApJ,Ferrand+2021ApJ,Polin+2022,Mandal+2023ApJ} suggest that anisotropies on small scales are dominated by RTI structures formed at the unstable contact surface between the shocked ejecta and the shocked CSM \citep{Chevalier+1992ApJ}, whereas features at larger scales are imprints of anisotropies internal to the SN ejecta. It is also suggested that inhomogeneities or clumps in the CSM do not imprint upon the RTI structures unless they are extremely massive. \cite{Warren+2013MNRAS} and \cite{Polin+2022} demonstrate that there exists a typical angular scale $\theta_0$ that separates anisotropies endemic to the SN explosion itself from anisotropies from turbulent activity at the shock interface. In the absence of strongly asymmetric large scale features such as a jet or large $^{56}$Ni clumps, this typical scale $\theta_0$ becomes the dominant angular scale, where most of the turbulent power resides. \cite{Warren+2013MNRAS} showed that this angular scale is a function of the age of the SNR, and suggested that this scale may be related to the size of fleece-like structures seen in several Type Ia SNRs, one of the most notable being Tycho's SNR. \cite{Polin+2022} and \cite{Mandal+2023ApJ} further show using their self-similar models that the dominant angular scale $\theta_0$ may be used to diagnose the density profile of the outermost layers of the SN ejecta. Their analysis requires the use of angular power spectral analysis on 3D distributions of density or radial velocity of the SNR. While this method is very powerful in determining properties of turbulence at all scales, its application is limited by the availability of data on the three dimensional structure of observed SNRs. Mapping out the full 3D structure of a SNR involves careful measurement of position and spectra of the SNR features, and extracting the velocity of these features from their spectra. This is an involved process and has only been performed for a handful of SNRs, e.g., 1E 0102.2-7219 \citep{Vogt+2010ApJ}, N132D \citep{Vogt+2011ApJ, Law+2020ApJ}, Tycho's SNR \citep{Millard+2022ApJ,Uchida+2024arXiv}, Cassiopeia A \citep{DeLaney+2010ApJ,Milisavljevic+2013,Milisavljevic+2015}, 0540-69.3 \citep{Larsson+2021ApJ} and the Crab Nebula \citep{Charlebois+2010AJ,Martin+2021MNRAS}.

Images of SNRs as extended sources are comparatively easier to obtain, especially since the advent of facilities like the Chandra X-ray Observatory and the James Webb Space Telescope. But these images cannot provide full information on the 3D structure of the remnants and suffer from projection effects. Thus, it's highly beneficial to develop a tool that can infer information on SNR structures from high-resolution images which may be compared to numerical SNR models. A common way to do so is to take a Fourier transform of the image \citep{Warren+2005ApJ}, but unfortunately the resulting power spectrum is typically washed out into a single power-law, even for systems where we can tell by eye there is a clear dominant mode, such as Tycho's SNR. A major step forward in this direction is the use of a wavelet transform \citep{Lopez+2009ApJ}, which was successfully employed to study the x-ray substructure of many SNRs \citep{Lopez+2011ApJ}. The idea is to convolve the SNR image with a Ricker wavelet, which can pick out power in structures of any given size. In this work, we expand on this idea to extract the power spectrum of images of SNRs. We use 3D numerical models to interpret and provide predictions for power spectrum of SNRs. Our SNR models start from a 1D carbon deflagration SN model,  and are used to generate synthetic thermal x-ray images. Power spectra obtained from these synthetic images are compared against angular power spectra obtained directly from the 3D models (as in \cite{Polin+2022} and \cite{Mandal+2023ApJ}) to ensure correct interpretation of the former. As a test case, we use this technique to measure the power spectrum of Tycho's remnant to determine the typical size of its fleece-like structures, as well as the nature of turbulent activity at small scales.

This paper is organized as follows. The numerical method and initial conditions are discussed in Section \ref{sec:numerical}. The technique for extracting power spectra from images is described briefly in Section \ref{sec:img_anly}. The results from power spectral analysis of the models and the synthetic images generated from them are discussed in Section \ref{sec:results}. The analysis of the power spectrum of Tycho's remnant is presented in Section \ref{sec:observations}. A discussion and a summary of our results are provided in Sections \ref{sec:discussion} and \ref{sec:conclusion} respectively.


\begin{figure*}
\centering
\gridline{\fig{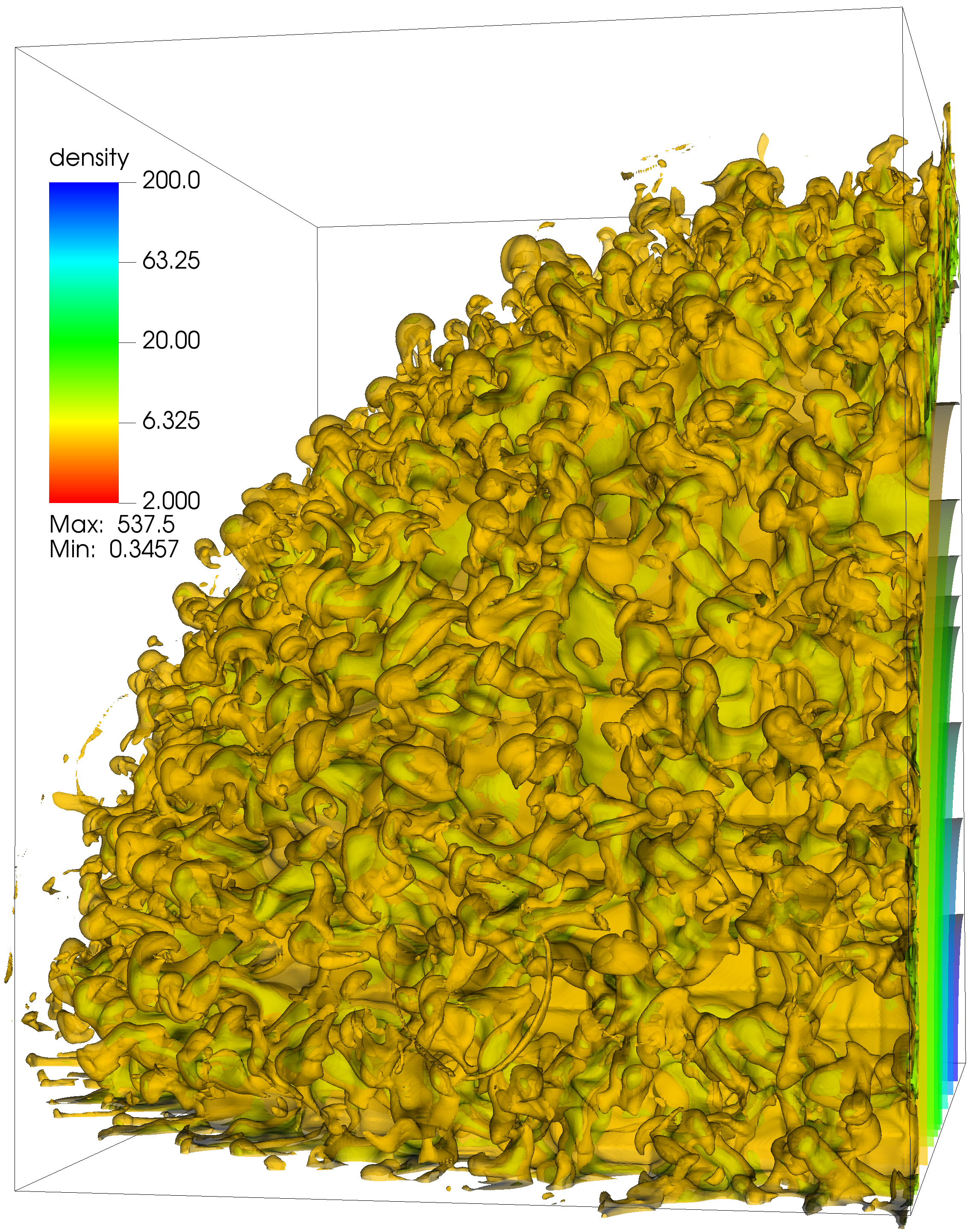}{0.4\textwidth}{}
          \fig{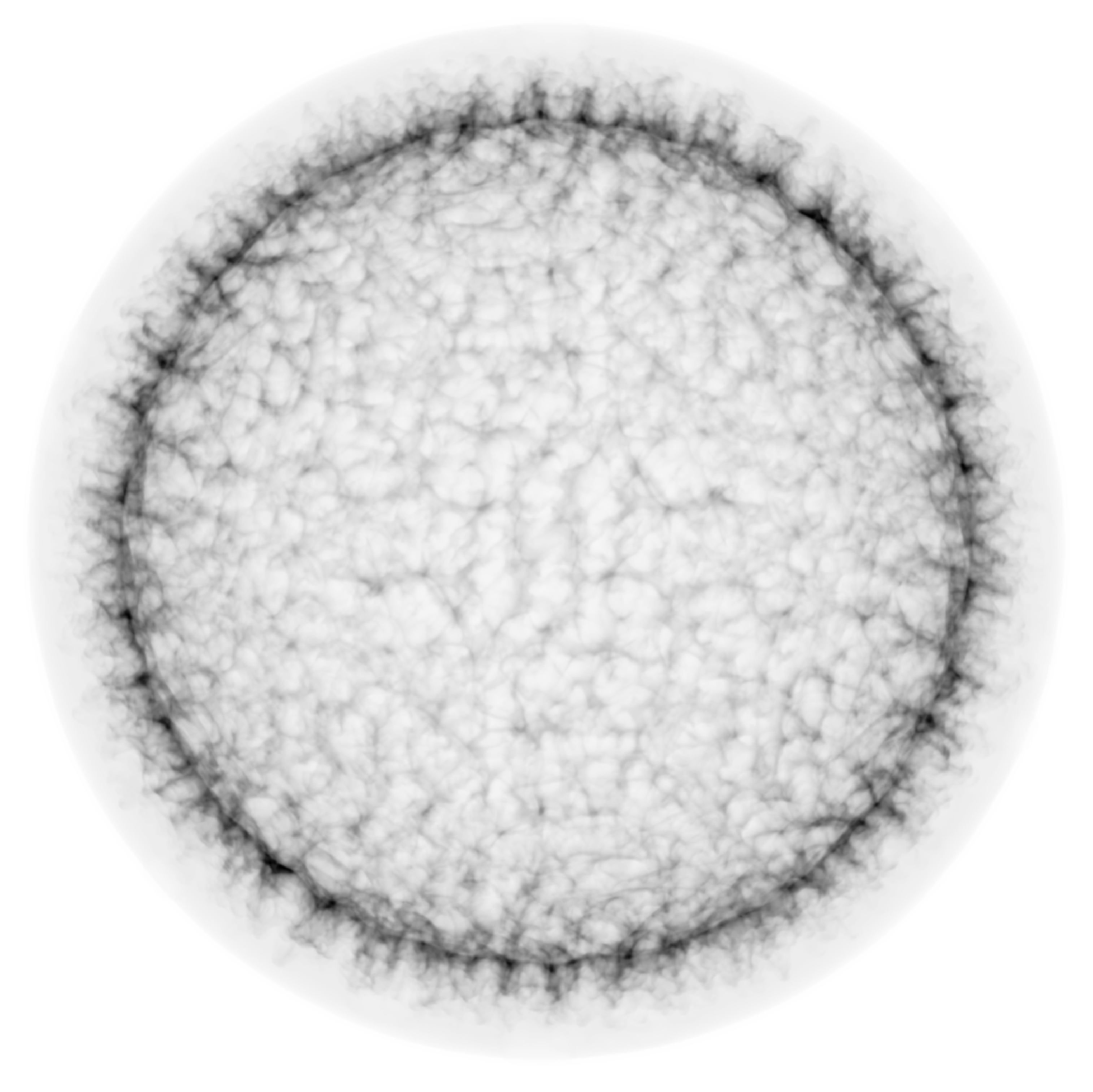}{0.48\textwidth}{}
          }
\caption{(\textit{Left}) Density isosurfaces in the shocked region of our fiducial model at $t/T=0.1$, where $t$ is the age of the remnant, and $T$ is the characteristic timescale when mass swept-up by the expanding shock is equal to the ejecta mass. (\textit{Right}) Square of density of the shocked plasma along parallel lines of sight, used as proxy for x-ray images. We assume the ejecta is optically thin and is ionized and heated to over $10^6K$ in the shocked region.}
\label{fig:first_look}
\end{figure*}

\section{Numerical method}  \label{sec:numerical}

Numerical calculations were performed using \sprout, a second-order expanding mesh hydro code \citep{Mandal+2023_sprout}. \sprout\, solves the equations of ideal, non-relativistic hydrodynamics on a Cartesian mesh:

\begin{equation}
\label{eq:euler}
    \begin{gathered}
        \partial_t(\rho) + \nabla \cdot ( \rho \mathbf{v} ) = 0 \\
        \partial_t( \rho \mathbf{v} ) + \nabla \cdot ( \rho \mathbf{vv} + P \overleftrightarrow{I} ) = 0 \\
        \partial_t\left( \frac{1}{2}\rho v^2 + \epsilon \right) + \nabla \cdot \left( \left( \frac{1}{2}\rho v^2 + \epsilon + P \right)\mathbf{v} \right) = 0,
    \end{gathered}
\end{equation}

where $\rho$, $\mathbf{v}$, $P$ and $\epsilon$ are density, velocity, pressure and internal energy of the fluid respectively. \sprout\, employs the moving mesh methodology \citep{Springel2010MNRAS,Duffell+2011ApJS} to expand its mesh with time, tracking the expansion of the remnant. No source terms were required for this problem. $P$ is related to $\epsilon$ via an adiabatic equation of state:

\begin{equation}
    P = (\gamma-1)\epsilon,
\end{equation}

where $\gamma$ is the adiabatic index. Most of the results presented here are based on a model with $\gamma=5/3$. In addition, we also solve for models with reduced values of the adiabatic index ($\gamma=3/2,\,4/3$). This has the net effect of cooling the shock and increasing the gas compressibility, mimicking the effect of cosmic ray acceleration by the shock front \citep{Blondin+2001ApJ}.

\subsection{Initial conditions} \label{subsec:initial}

We consider a typical thermonuclear or Type Ia SN for modeling our ejecta. \cite{Dwarkadas+1998ApJ} showed that an exponential profile most closely matches the ejecta density profile of the numerical W7 model for a thermonuclear SN experiencing deflagration \citep{Nomoto+1984ApJ}. We therefore assume an exponential density profile for the SN ejecta, which initially expands homologously ($v=r/t$) into the stationary CSM, assumed to have a constant density. Thus, our initial conditions are as follows:

\begin{equation}
   \begin{gathered}
       \mathrm{Ejecta:}\;\;\rho \propto t^{-3}e^{-v/v_e},\;\mathbf{v}=\mathbf{r}/t \\
       \mathrm{CSM:}\;\;\rho = \mathrm{constant},\;\mathbf{v}=0 \\
       P = 10^{-6}\rho\;\mathrm{(both\;cases).}
   \end{gathered}
\end{equation}

The normalization for the ejecta density profile and the characteristic ejecta velocity $v_e$ are set by the ejecta mass $M$ and explosion energy $E$. Now we note that $v_e$ scales as

\vspace{-2mm}

\begin{equation}
\label{scl:v}
    v_e\sim(E/M)^{1/2},
\end{equation}

and that this gives us a scaling for the radius $r$ of the remnant at time $t$:

\vspace{-4mm}

\begin{equation}
\label{scl:r}
    r \sim v_e t.
\end{equation}

The radius of the remnant at time $t$ provides a scaling for the CSM mass swept up by the forward shock till that time:

\vspace{-4mm}

\begin{equation}
\label{scl:M}
    M_{\mathrm{swept}} \sim r^3 \rho,
\end{equation}

where $\rho$ denotes the CSM density. Scalings (\ref{scl:v}-\ref{scl:M}) provide an estimate for the timescale when the mass of the ejecta $M$ equals the swept-up mass (also obtained from dimensional analysis):

\begin{equation}
\label{scl:T}
\begin{gathered}
    M \sim  r^3 \rho\; \mathrm{or,}\;r \sim (M/\rho)^{1/3} \sim v_e T , \\
    \mathrm{or,}\;T \sim (M/\rho)^{1/3}/v_e
\end{gathered}
\end{equation}

We use (\ref{scl:T}) to define a characteristic time $T$:

\begin{equation}
\label{eq:timescale}
    \begin{split}
        T &\equiv M^{5/6} E^{-1/2}\rho^{-1/3}  \\
        &= 563\left(\frac{M}{M_{ch}}\right)^{5/6} E_{51}^{-1/2}n_0^{-1/3}\mathrm{\;yrs},
    \end{split}
\end{equation}

where $M_{ch}=1.4M_{\odot}$ is the Chandrasekhar mass, $E_{51}=E/(10^{51}\mathrm{ergs}$), and $n_0 =\rho/(2.34\times10^{-24}\mathrm{g}$) is the number density of the circumstellar medium, assuming a $10:1$ H:He ratio. Note that this scaling is not unique. For example, \cite{Warren+2013MNRAS} choose a different characteristic age $T'$, which is related to our choice as:

\begin{equation}
\label{eq:T_WB}
    T' = 0.43T.
\end{equation}

Our calculations start at $t/T=10^{-4}$ and run till $t/T=10$. The models cover an octant of the full sphere and have a fiducial resolution of $(512)^3$.



\subsection{Power spectral analysis} \label{subsec:angular_ps_anly}

Following \cite{Polin+2022}, we define spherical surface distributions of radially integrated density and radial velocity as follows:

\begin{equation}
    \begin{split}
        \left<v_r\right>(\theta,\phi) = \frac{\int P(r,\theta,\phi)\mathbf{v}(r,\theta,\phi)\cdot \mathbf{dr}}{\int P(r,\theta,\phi)dr} \\
        \left<\rho\right>(\theta,\phi) = \frac{\int \rho(r,\theta,\phi) P(r,\theta,\phi)dr}{\int P(r,\theta,\phi)dr} \\
    \end{split}
\label{eq:surface_maps}
\end{equation}

The pressure weighting takes into account only contributions from the shocked, turbulent region, since the pressure is significant only in the shocked fluid. These distributions are then normalized by dividing with their angle-averaged value. The relative amplitude of anisotropies in, say the density distribution, may be obtained by expanding it as a sum of spherical harmonics:

\begin{equation}
    \left<\rho\right>(\theta,\phi) = \sum\limits_{l,m} a_{lm} Y_{lm}(\theta,\phi)
\end{equation}

The amplitudes of the harmonics are used to calculate a power spectrum as a function of the harmonic number $l$:

\begin{equation}
    C_l = \frac{1}{2l+1}\sum_{m=-l}^{+l}\left|a_{lm}\right|^2
\end{equation}

We obtain power spectra for our normalized surface maps using the SHTOOLS package \citep{SHTOOLS}. The raw power spectra were smoothened following the same procedure as \cite{Polin+2022}.

\section{Image analysis technique}  \label{sec:img_anly}


We generate proxies for x-ray images of SNRs by integrating the square of the density of the shocked plasma from our models. It is assumed that x-ray emission is dominated by thermal emission from plasma heated over a million Kelvin, and the remnant is optically thin. An example is shown in the right panel of Figure \ref{fig:first_look}, constructed from our fiducial model at $t/T=0.1$. The power at different length scales in these images may be calculated using a wavelet transform, as first done by \cite{Lopez+2009ApJ}. We choose a different implementation for this, namely the modified $\Delta$-variance technique \citep{Ossenkopf+2008A&A,Arevalo+2012MNRAS}. This technique  produces a low-resolution power spectrum that may be directly compared to the angular power spectrum produced using our 3D models. Additionally, the $\Delta$-variance method is designed to eliminate spurious modes from gaps in the data. Given an image $I(x,y)$ and a mask $M$ to hide gaps or undesired edges in the image, it's possible to calculate a filtered image that highlights only structures at a length-scale $\sigma$ (in pixel coordinates):


\begin{equation}
\label{eq:filtered_im}
    \Tilde{I}_{\sigma}(x,y) = \left( \frac{G_{\sigma_1}\ast(I\times M)}{G_{\sigma_1}\ast M} - \frac{G_{\sigma_2}\ast(I\times M)}{G_{\sigma_2}\ast M} \right) \times M
\end{equation}

where $G_{\sigma_1}$ and $G_{\sigma_2}$ are Gaussians with standard deviations $\sigma_1=\sigma/\sqrt{1+\epsilon}$ and $\sigma_2=\sigma\sqrt{1+\epsilon}$, respectively. The mask $M$ eliminates edge effects from the filter, as is found from Equation \ref{eq:filtered_im}. In our case, the mask is a disk of chosen radius centered on the SNR center. We set $\epsilon=10^{-3}$, following \cite{Arevalo+2012MNRAS}. The power $P$ at the length-scale $\sigma$ is proportional to the variance V of the filtered image \citep{Churazov+2012MNRAS,Arevalo+2012MNRAS}:

\begin{equation}
    V_{\sigma} = \int \Tilde{I}_{\sigma}^2 dxdy.
\end{equation}

We can relate the pixel length scale $\sigma$ to a spherical harmonic $l$ in the remnant, by noting that the angle subtended at the center by an arc of length $\sigma$ is of the order of the angular scale associated with the spherical harmonic $l$:

\vspace{-5mm}

\begin{equation}
\label{eq:sigma_to_l}
    \sigma/r \sim 1/l,
\end{equation}

where $r$ is the radius of the reverse shock in the image in pixel coordinates. The radius of the reverse shock is taken to be the relevant radius because that's roughly where the turbulent structures reside. We find $l\approx 1.6r/\sigma$ after calibration with our dataset. Finally, the variance at an angular harmonic $l$ can be related to the power at that scale through a scale-dependent normalization \citep[Eq. A8 of][]{Arevalo+2012MNRAS}\footnote{although that equation uses the wavenumber $k(=0.225/\sigma)$, it holds for our case because $l$ and $k$ are both inversely proportional to $\sigma$.}:

\begin{equation}
    P(l) \propto l^{-2}V_l = l^{-2} \int \Tilde{I}_l^2\, dxdy
\end{equation}

While extracting power spectrum from a SNR image, it is beneficial to subtract the azimuthal average of the image from itself. This eliminates large scale projection-based features such as limb brightening (such as the bright ring in the example image in Figure \ref{fig:first_look}), but can potentially tamper with real large scale features (see bottom right panel of Figure 1 in \cite{Warren+2005ApJ}, for example). Hence, we consider both the example image and and its azimuthal average subtracted version while testing the $\Delta$-variance technique for our problem. In addition, we  apply masks (centered on the SNR center) with different radii so that the smallest ones ignore large scale projection effects.

\begin{figure}
\centering
\includegraphics[width=0.45\textwidth]{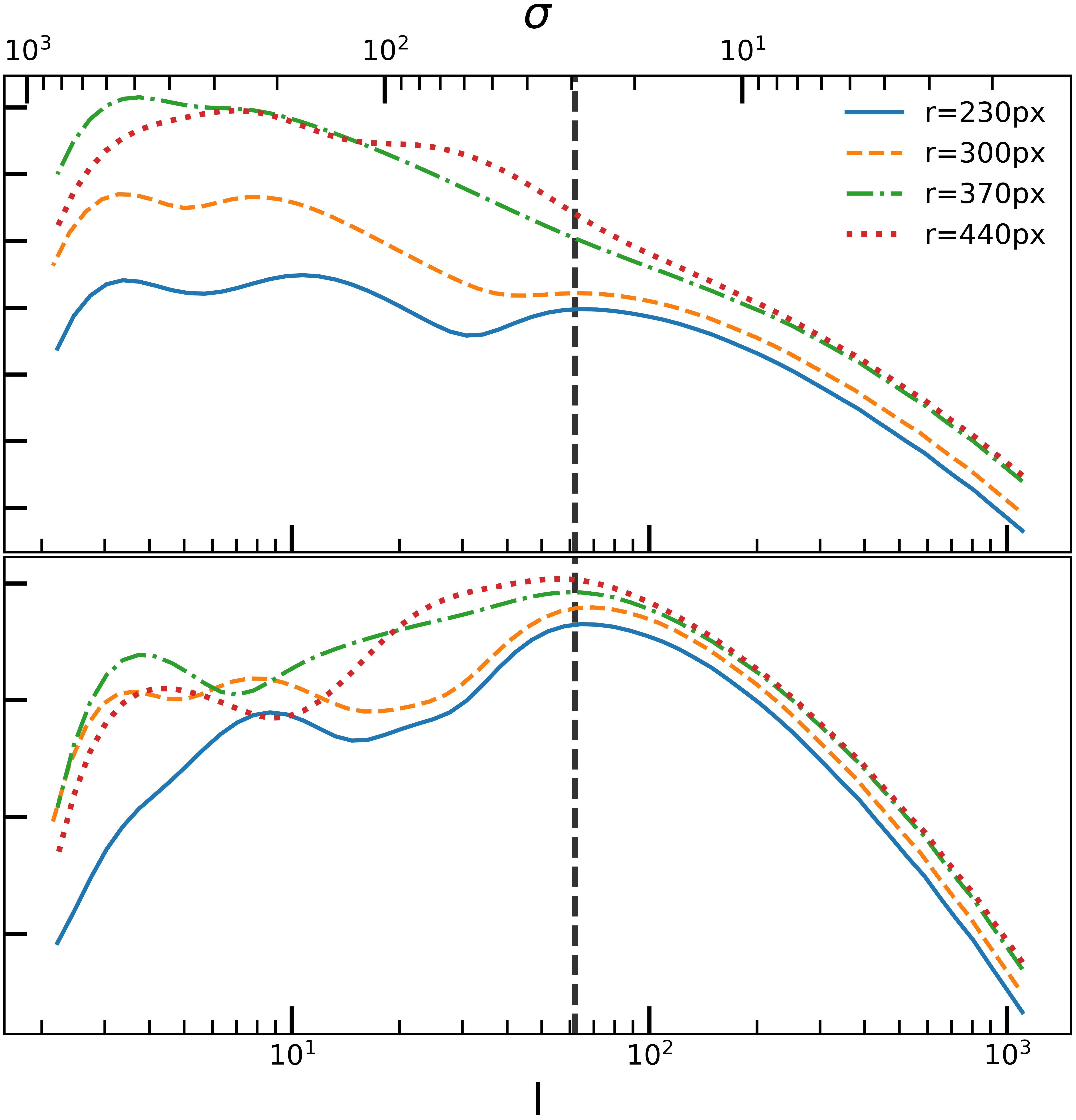}
\vspace{-2mm}
\caption{Power spectra of the synthetic image in right panel of Figure \ref{fig:first_look}, made from the raw image (top) and its azimuthal average subtracted version (bottom). For each image, four power spectra were obtained (corresponding to the four curves) using masks of different radii. The radii of the forward shock and the reverse shock are $\sim360$ and $\sim450$ pixels, respectively. The smaller masks ignore most of large scale projection effects and help pick out small scale features such as in Tycho's remnant.}
\label{fig:img_anly}
\end{figure}

Figure \ref{fig:img_anly} shows the resulting power spectra for the raw image and azimuthal average subtracted image on the upper and lower panels respectively. The power spectra for the raw image with the two largest masks (red dotted and green dash-dotted curves in the top half of the panel) decrease almost monotonically with increasing wavenumber. As the mask size is decreased (the blue solid and orange dashed curves), the power at large scales or small wavenumbers decreases and a second peak emerges. The smaller masks ignore large scale projection effects, like the bright ring in the SNR image, and thus admit lesser power at large scales. The power spectra for the azimuthal average subtracted images (lower panel) converge to the same value of the peak harmonic, and the same shape at small scales (corresponding to a power law $P(l)\propto l^{-3}$), irrespective of the mask size. These values also coincide with those from the power spectra of the raw image with the smallest mask (blue curve in the upper panel), demonstrating convergence to the correct answer for the azimuthal average subtracted images.

\begin{figure}
\centering
\includegraphics[width=0.48\textwidth]{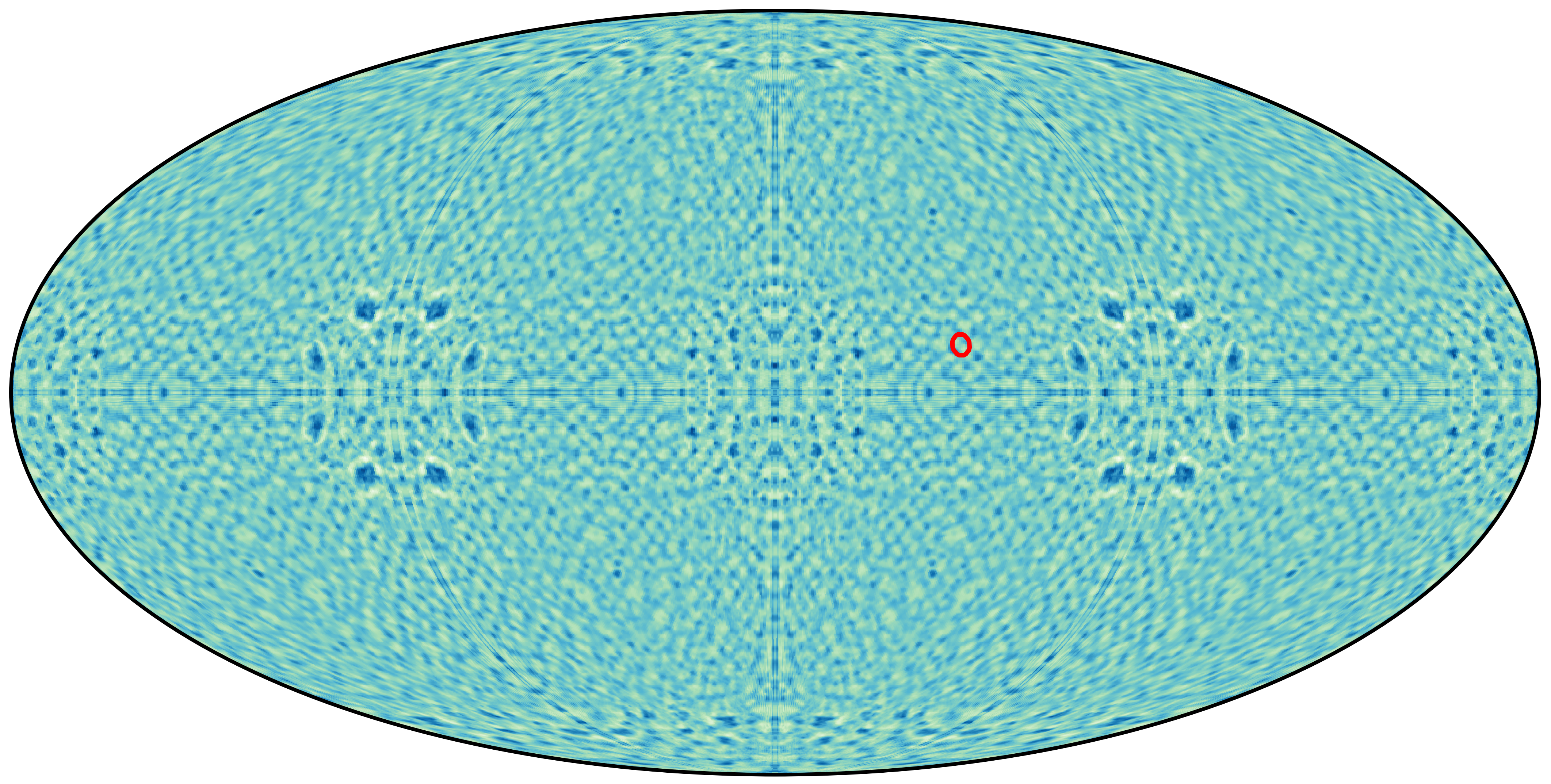}
\includegraphics[width=0.48\textwidth]{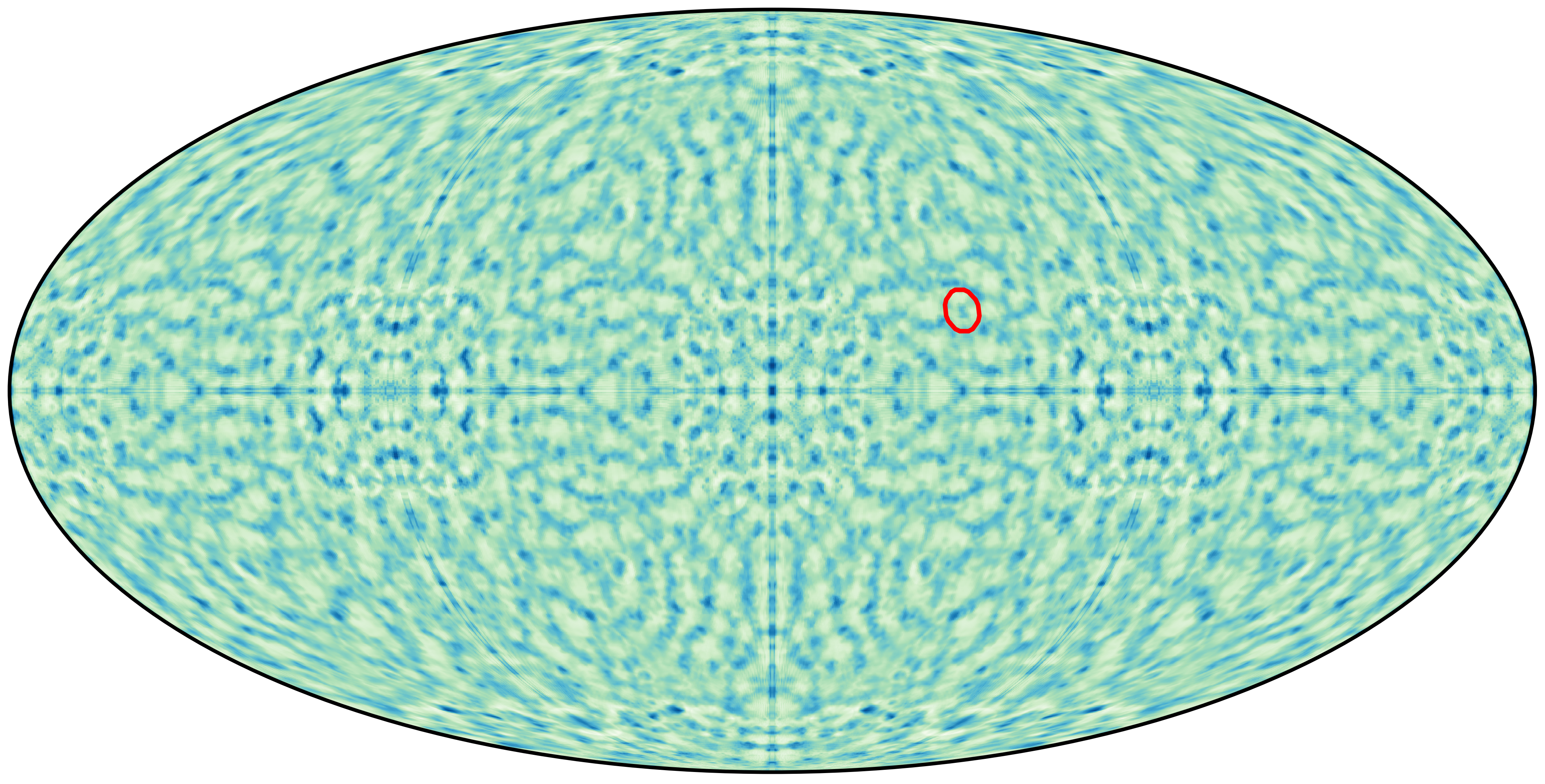}
\includegraphics[width=0.48\textwidth]{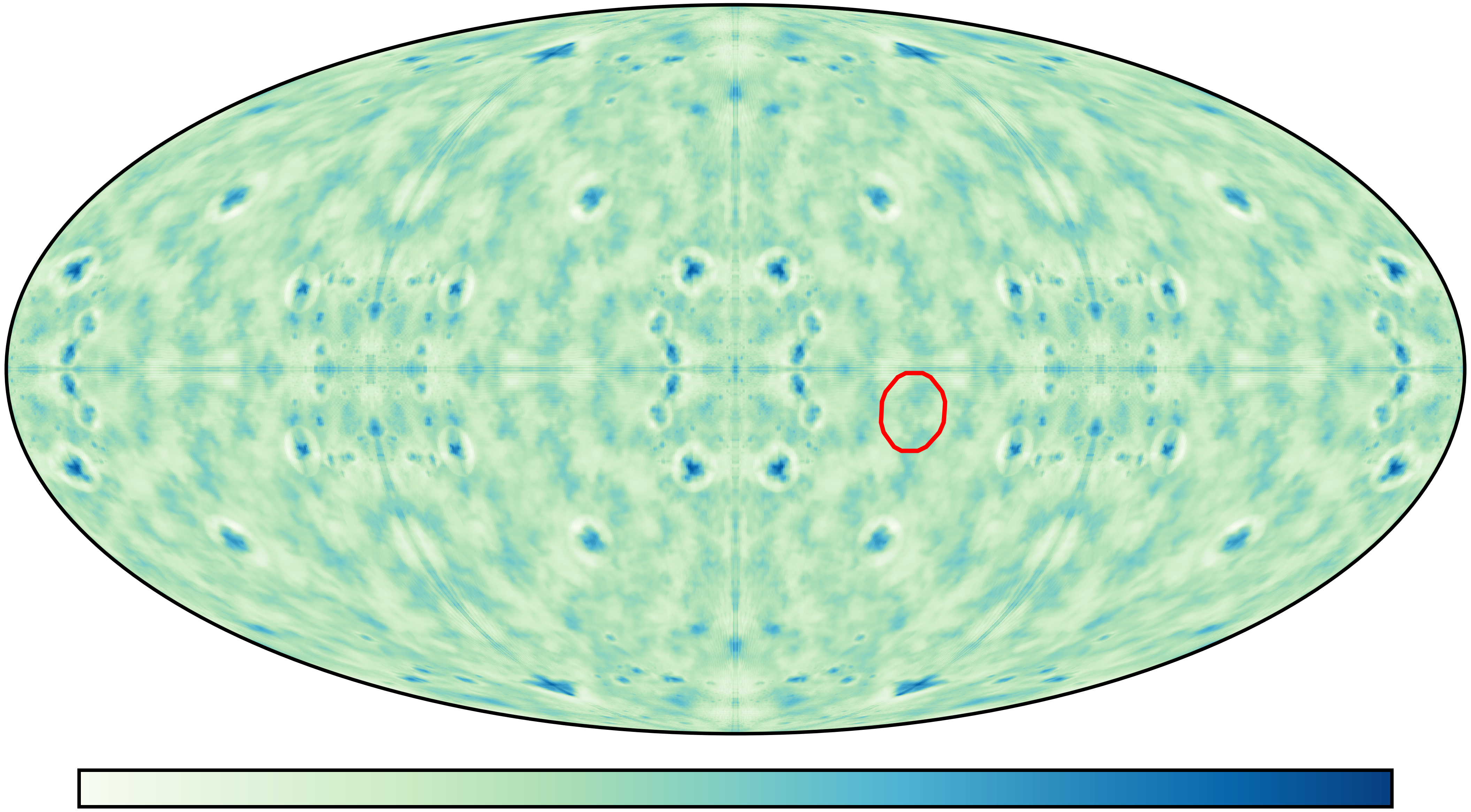}
\caption{Mollweide projections of the radially integrated column density of the shocked plasma in our fiducial model, plotted at three different time instants: \textit{(top)} $t/T=0.01$, \textit{(middle)} $t/T=0.1$, and \textit{(bottom)} $t/T=1.0$. The overlaid red circles mark the approximate size of the dominant angular scale at each time instant.}
\label{fig:mollweide}
\end{figure}

\begin{figure*}
\centering
\gridline{\fig{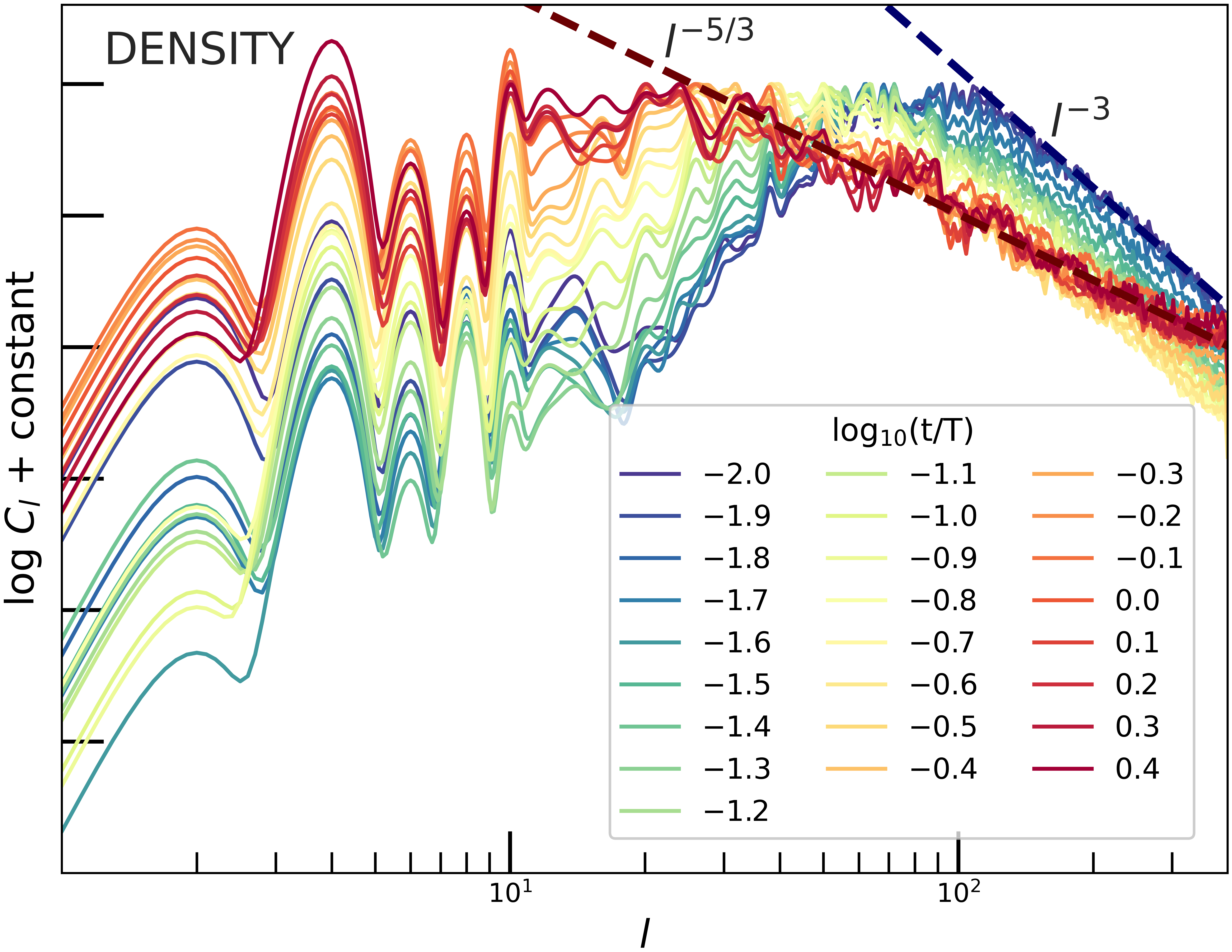}{0.5025\textwidth}{}\fig{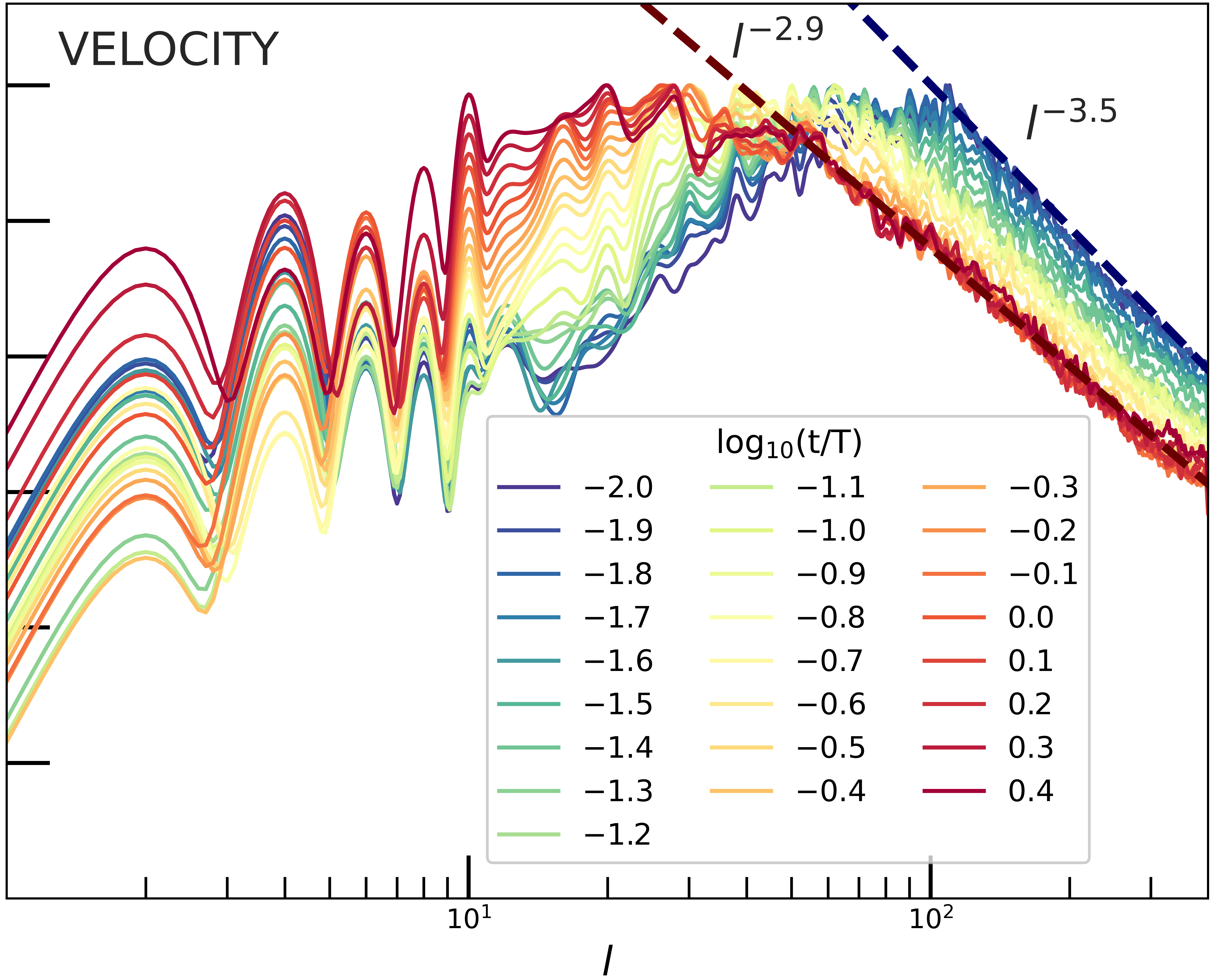}{0.48\textwidth}{}}
\vspace{-8mm}
\caption{\textit{(Left)} Density and \textit{(Right)} velocity power spectrum for the fiducial model ($\gamma=5/3$) as a function of the scaled age $t/T$ (defined in Section \ref{subsec:initial}). The velocity power spectra at small scales (large values of angular harmonic $l$) have a steep shape, with signs of shallowing with time. The density power spectra at small scales have a similar steep slope at early times, but switch to a Kolmogorov turbulence spectrum ($C_l \propto l^{-5/3}$) at late times, around $t/T=1$.}
\label{fig:fiducial_spectra}
\end{figure*}

\section{Results}   \label{sec:results}

\subsection{Hydrodynamics}  \label{subsec:hydro}

The expansion of the ejecta against the CSM generates RTI structures as shown in the left panel of Figure \ref{fig:first_look}. The Mollweide projections of the radially integrated density map (as defined by Equation \ref{eq:surface_maps}) at three different time instants in Figure \ref{fig:mollweide} show that the angular scale of the strongest RTI structures increases in size as time progresses, as shown previously by \cite{Warren+2013MNRAS}. Figure \ref{fig:fiducial_spectra} shows the power spectra constructed from these density maps and the velocity maps as a function of the scaled age ($t/T$) of our remnant models. Each power spectrum is found to have the shape of a broken power-law, which peaks at some wavenumber we call $l_0$. The peak or break wavenumber represents the angular scale of the strongest turbulent structures in the remnant. As expected from Figure \ref{fig:mollweide}, $l_0$ decreases (or the associated angular scale increases) with the age of the remnant. To compare with previous works, we define the instantaneous or `effective' power-law slope of the ejecta density profile encountering the reverse shock as:

\begin{equation}
\label{eq:n_eff}
    n_{\mathrm{eff}} \equiv \left| \frac{d\,\mathrm{log}\,\rho}{d\,\mathrm{log}\,r}\right|_{r=r_{RS}} = \frac{r_{RS}}{v_e t},
\end{equation}

where the last equality follows from our choice of the exponential profile for the ejecta density. We find that our models follow the relation  $l_0\sim 10n_{\mathrm{eff}}$. \cite{Polin+2022} point out that the results by \cite{Warren+2013MNRAS} are also consistent with this relation. This proportionality is also found by \cite{Polin+2022} and \cite{Mandal+2023ApJ} for their self-similar models with a power-law ejecta ($\rho\propto r^{-n}$) and a range of values of $n$.

Figure \ref{fig:fiducial_spectra} shows that both the density and velocity power spectra have a steep slope ($C_l\propto l^{-3.5}$) for $l>l_0$ at relatively early times ($t/T=10^{-2}$). As time progresses, the density spectrum becomes shallower ($C_l\propto l^{-5/3}$). Surprisingly, this is the power law expected for a turbulent cascade \citep{Kolmogorov1941DoSSR}, but hasn't been observed in turbulent activity in numerical SNR models so far to our knowledge.

\subsection{Effect of cooling}  \label{subsec:cooling}

\begin{figure*}
\centering
\gridline{\fig{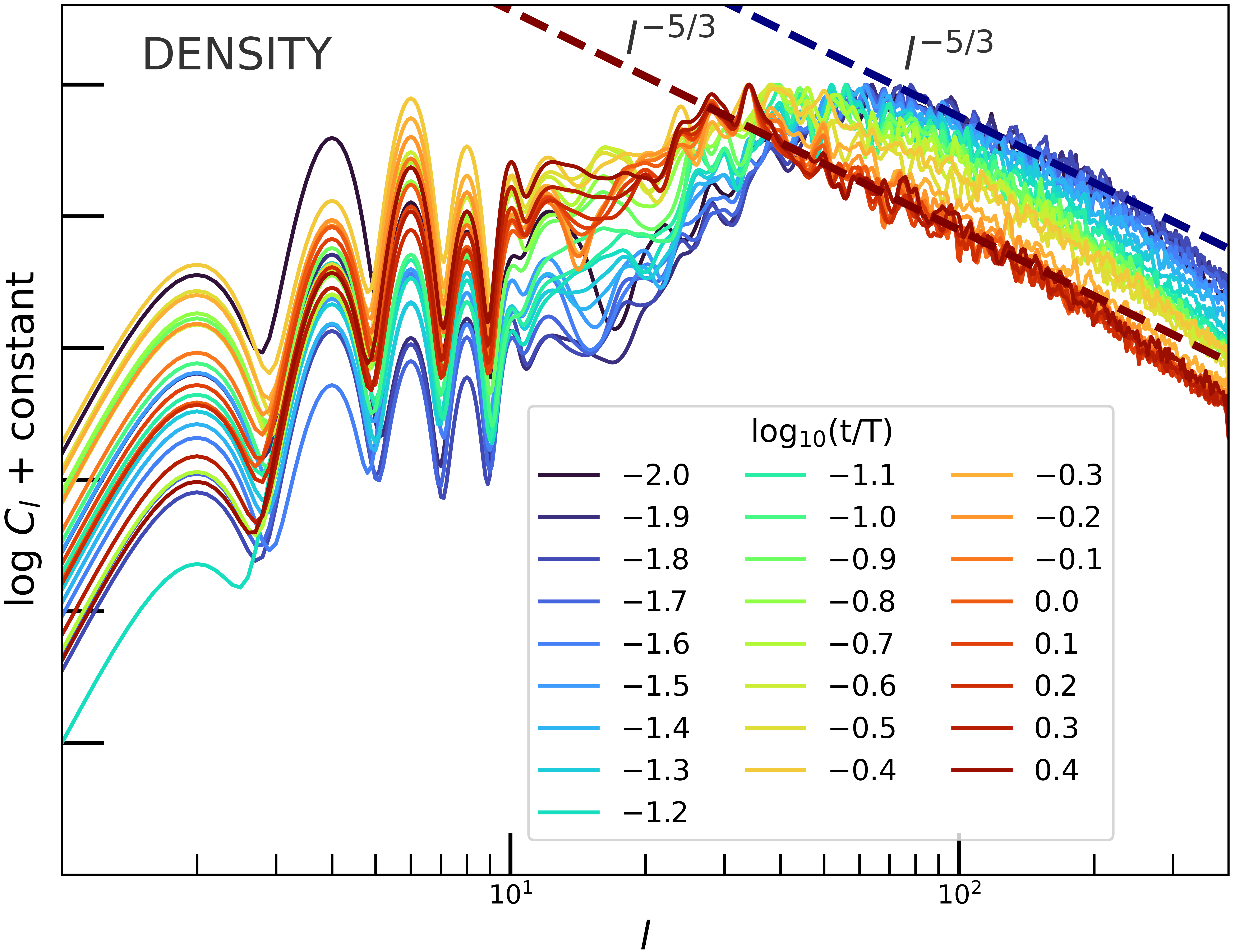}{0.503\textwidth}{}
          \fig{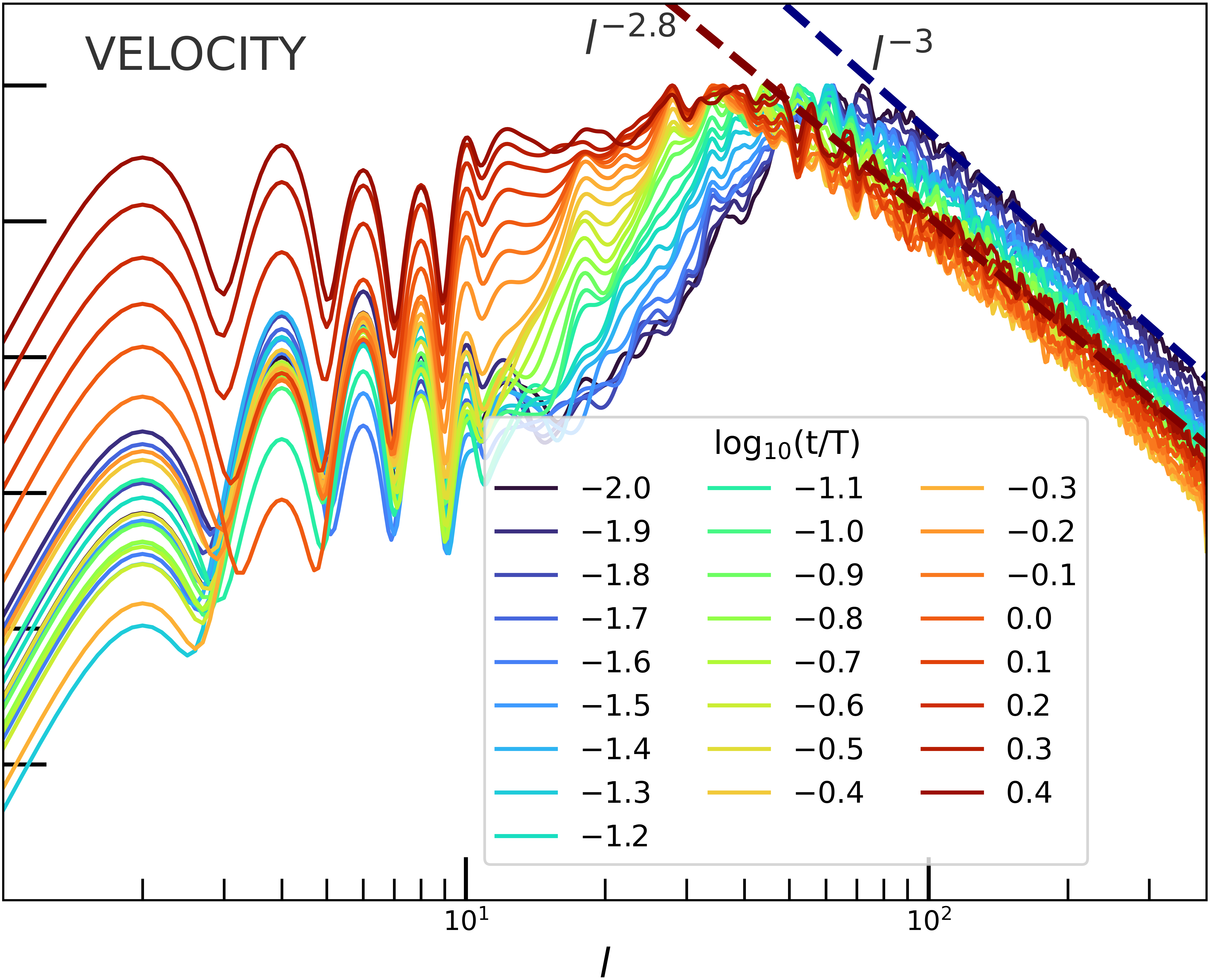}{0.48\textwidth}{}
          }
\vspace{-8mm}
\caption{The same as Figure \ref{fig:fiducial_spectra} but for $\gamma=4/3$, to mimick shock cooling via particle acceleration. The velocity power spectra behave similarly to their fiducial counterpart, albeit with a relatively shallower slope. In contrast, the density power spectra have a Kolmogorov-like slope ($C_l \propto l^{-5/3}$) at small scales even at early times. At very large values of $l(>100)$, the spectra steepen to a $\sim l^{-3}$ shape.}
\label{fig:cooling_spectra}
\end{figure*}

It has been argued for long that shocks in SNRs are the sites of cosmic ray acceleration. Accelerated particles transport thermal energy away from the shock, cooling it down, and increasing the compressibility of the shocked gas. \cite{Blondin+2001ApJ} demonstrate that the effects of particle acceleration on the shocked region may be reproduced qualitatively by lowering the adiabatic index. This results in the narrowing of the shocked region and possible contact of the RTI fingers with the forward shock or even unshocked CSM ahead of it, as also found by \cite{Warren+2013MNRAS}. Our models with reduced values of $\gamma$ ($3/2$ and $4/3$) exhibit similar features. The power spectra for the $\gamma=4/3$ model are shown in Figure \ref{fig:cooling_spectra}. They have the same overall shape of a broken power law as before. The shape of the density power spectra at $l>l_0$ however corresponds to that for a turbulent cascade ($C_l \propto l^{-5/3}$) at all times. This hints at an earlier onset of turbulent cascade as compared to the $\gamma=5/3$ models. Additionally, the density power spectra steepen to $C_l \propto l^{-3}$ for some large value of $l$ ($\sim100$ for $t/T=2.5$; see blue curve in the left panel of Figure \ref{fig:cooling_spectra}). In contrast, the velocity power spectra behave similarly to those for the $\gamma=5/3$ models. The other significant difference between the models with different adiabatic indices is the dominant angular scale. The corresponding values of $l_0$ are smaller for models where shock cooling is taken into account. 

Thus, the angular power spectrum of SNRs that haven't experienced significant perturbative activity should have a broken power-law shape whose specifics depend on the remnant age and the efficiency of shock cooling due to particle acceleration. In the next section, we address the question of whether power spectra extracted from images of SNRs are sensitive to these details.



\subsection{Expectation for SNR observations}  \label{subsec:img_anly}

\begin{figure*}
\centering
\includegraphics[width=0.95\textwidth]{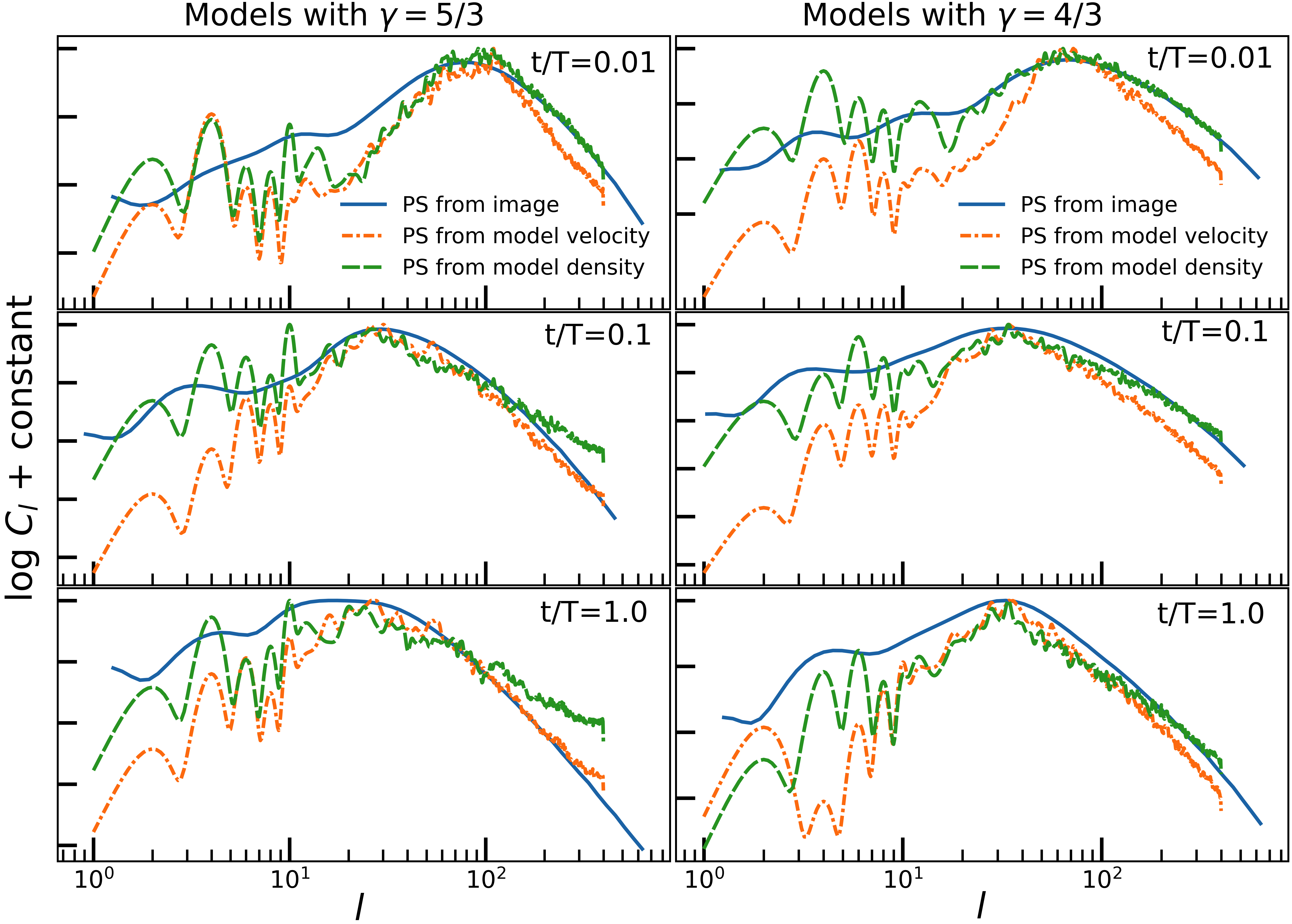}
\vspace{-2mm}
\caption{Power spectra of proxy x-ray images from SNR models (solid curves), compared against angular power spectra of density and velocity taken directly from the corresponding models with $\gamma=5/3$ (left) and $\gamma=4/3$ (right). Three different time instants are depicted: $t/T=0.01$ (top), $0.1$ (middle), and $1.0$ (bottom). The image power spectra correctly picks out the dominant angular harmonic (from angular power spectra) in all cases, demonstrating accuracy of the $\Delta$-variance technique. At large values of $l$, the image power spectra match the density spectra at early times, but the velocity spectra at late times.}
\label{fig:im_spectra_all_models}
\end{figure*}

\begin{figure}
\centering
\includegraphics[width=0.48\textwidth]{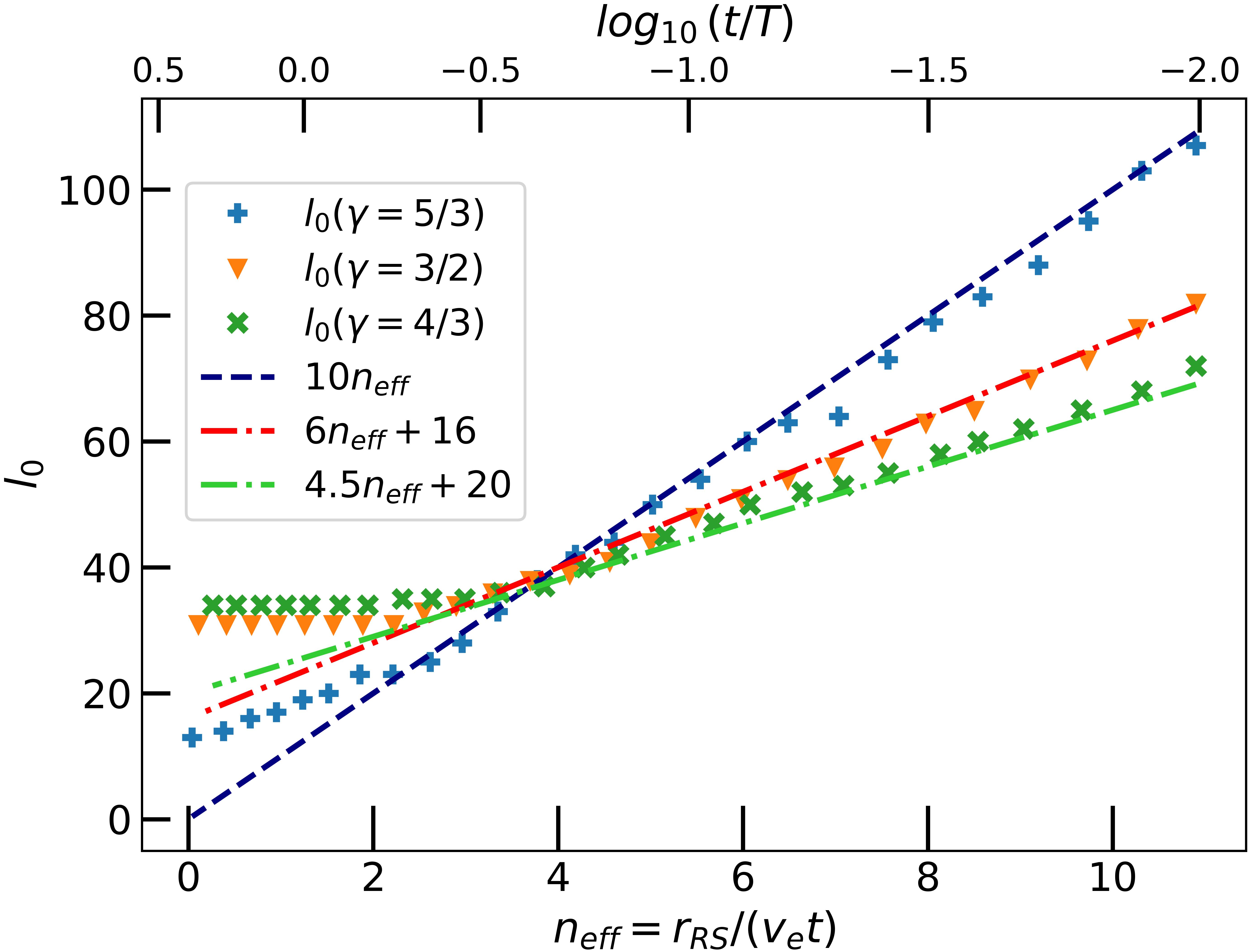}
\vspace{-6mm}
\caption{The dominant angular harmonic ($l_0$) as obtained from the image power spectra, vs the effective power-law slope $n_{\mathrm{eff}}$ (defined in Equation \ref{eq:n_eff}) of the outermost ejecta. For all three models ($\gamma=5/3$, $3/2$, and $4/3$), the relation between $l_0$ and $n_{\mathrm{eff}}$ is linear, with the slope decreasing with $\gamma$. This relation seems to break for low values of $n_{\mathrm{eff}}$, where $l_0$ saturates at some minimum value (seen most clearly for ($\gamma=3/2$, and $4/3$).}
\label{fig:l0_vs_neff}
\end{figure}

Here we compare the power spectrum of the proxy images to the angular power spectra of the corresponding 3D models, as plotted in Figure \ref{fig:im_spectra_all_models} for the $\gamma=5/3$ and $\gamma=4/3$ models at three time instants (on the left and the right panels, respectively). In all cases, the power spectra of images picks out the same break harmonic $l_0$ as the angular power spectra. But the exact relationship between $l_0$ and $n_{\mathrm{eff}}$ is dependent on the adiabatic index of the model (or the amount of shock cooling experienced by the system), as seen in Figure \ref{fig:l0_vs_neff}. We also note that the shape of the image power spectra at large wavenumbers (or small length scales) agrees with the angular power spectra of density at relatively early times and switches to match the velocity power spectrum instead at late times. Despite this unexpected behavior, this data gives us a prediction for power spectra of observed SNRs, if their scaled age (as defined in Section \ref{subsec:initial}) and an effective adiabatic index can be estimated. Taking cue from Figure \ref{fig:im_spectra_all_models}, we estimate the power spectrum of surface brightness fluctuations of SNR images as a broken power law:

\begin{equation}
    P_l \propto \frac{1}{(l/l_0)^{-n_1}+(l/l_0)^{-n_2}},
\end{equation}

where $n_1(\approx3)$ and $n_2(\approx-3)$ are power-law slopes of the small wavenumber and large wavenumber parts of the image power spectrum, respectively. The quantities $l_0$, $n_1$ and $n_2$ completely specify the shape of the power spectrum and are solely functions of the scaled age of the remnant, which is as defined in Equation \ref{eq:timescale}. We have made available a \texttt{Python} script\footnote{available for download \hyperlink{https://github.com/smandal97/SNR_power_spectral_analysis}{here}} that takes the scaled age ($\tau$) of a remnant and the adiabatic index ($\gamma$) as inputs and produces the predicted power spectrum of its image. As will be shown in the next section, the comparison between the predicted and the observed power spectrum of a remnant may be used to infer properties of the explosion and the surrounding medium.

\begin{figure*}
\centering
\gridline{\fig{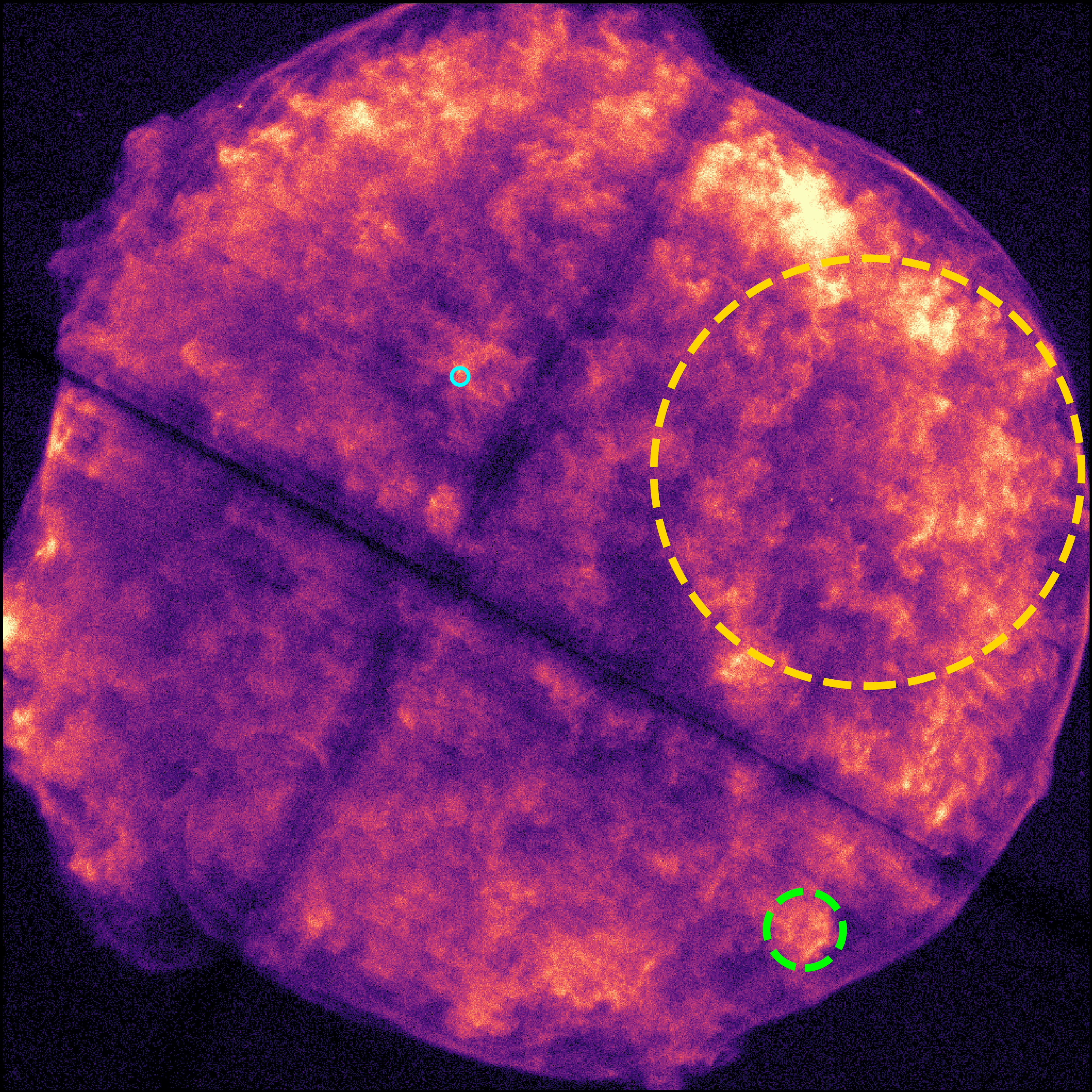}{0.467\textwidth}{}
          \fig{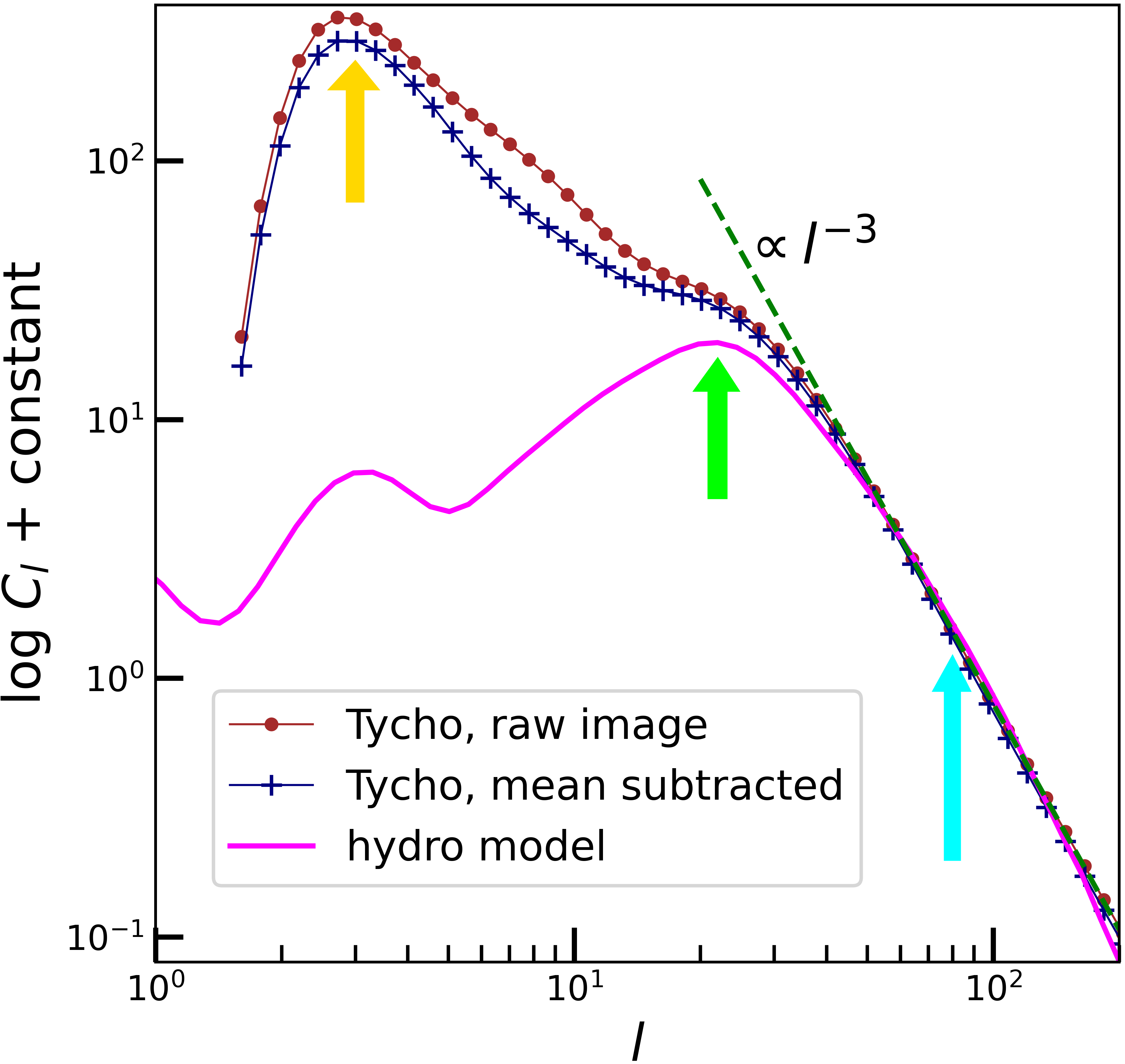}{0.493\textwidth}{}
          }
\vspace{-6mm}
\caption{(Left) Broadband x-ray image of Tycho's remnant \citep{Warren+2005ApJ} and (right) its power spectra obtained using the $\Delta$-variance technique. The image is overlaid with circles of sizes corresponding to harmonics marked (with arrows of respective colors) on the spectra. The green circle is roughly the size of fleece-like structures in Tycho, confirming the break harmonic in Tycho is indeed the dominant angular harmonic we seek. Tycho's power spectrum is compared against image power spectrum of a numerical SNR model with a peak harmonic coinciding with Tycho's break harmonic (corresponding to $\gamma=5/3$ and $t/T=0.8$). The power spectrum of Tycho and that of the model match very well at small scales ($C_l \propto l^{-3}$ at large $l$), but Tycho is seen to have additional power at large scales (small $l$), pointing to anisotropies endemic to the supernova explosion. }
\label{fig:tycho_anly}
\end{figure*}

\section{Application to observations}
\label{sec:observations}

We now apply the $\Delta$-variance technique as described in Section \ref{sec:img_anly} to an x-ray image of Tycho's SNR obtained by the \textit{Chandra} X-ray Observatory \citep{Warren+2005ApJ}, shown in the left panel of Figure \ref{fig:tycho_anly}. The power spectra from the raw image and the azimuthal average subtracted image are shown on the right panel. The two spectra are very similar, indicating the absence of strong limb brightening effects. There is no clear peak, but a change of slope at $l\approx23$. This break is reminiscent of some of the spectra in the upper right panel of Figure \ref{fig:img_anly} and leads to the speculation that the break harmonic indeed corresponds to the fleece-like structures seen in Tycho. To investigate this further, the image of Tycho is overlaid with circles of three different diameters, corresponding to three angular harmonics marked (with arrows of the corresponding color) on the power spectrum. We find that the green circle, corresponding to $l\approx23$ (or a diameter of $\sigma\approx40$ pixels), correctly picks out the size of the fleece-like structures. \cite{Lopez+2011ApJ} obtain a similar size for the fleece-like structures in Tycho's SNR. Their estimate of the size is about $9\%$ of the radius of the x-ray remnant, which closesly matches our result ($\sim8\%$). In addition, there is a large scale anisotropic distribution of brightness fluctuations towards the northwestern part of Tycho (marked roughly by the yellow circle) that likely corresponds to the low-$l$ modes of the obtained power spectrum.

\subsection{Comparison of Tycho's remnant to our models}

To compare Tycho's power spectrum against that of our model, we note that $l_0\approx23$ corresponds to a scaled age $t/T\approx0.5$, using Figure \ref{fig:l0_vs_neff} (see top x-axis). Note that $l_0\approx23$ confines us to the choice $\gamma=5/3$ since $l_0$ saturates at a minimum value of $\sim30$ for $\gamma=4/3$ or $\gamma=3/2$. Therefore, the image power spectrum for our SNR model corresponding to $\gamma=5/3$ and $t/T=0.5$ has a peak harmonic $l_0$ that coincides with Tycho's break harmonic ($l\approx23$). This is plotted in the right panel of Figure \ref{fig:tycho_anly} (solid curve). The shape of the model power spectrum matches that for Tycho near perfectly at small scales, demonstrating the accuracy of small-scale turbulence description in our models. But Tycho's power spectrum shows additional power on larger scales that cannot be explained by allowing for projection effects on our model alone. Prediction of the scaled age of Tycho's remnant additionally gives an independent estimate of the surrounding average CSM density, knowing Tycho has an age of 433 years at the time of observation (the SN event was observed in 1572). Using this in $t/T=0.5$ (with T as in Equation \ref{eq:timescale}) gives a number density $n_0\approx0.28(M/M_{ch})^{5/2}E_{51}^{-3/2}\;\mathrm{cm^{-3}}$ for the CSM. Figure \ref{fig:l0_vs_neff} also tells us that $n_{\mathrm{eff}}\sim2$ for $l_0\approx23$,  that is, the reverse shock in Tycho's SNR is running into a portion of the ejecta whose density profile is quite shallow and can be approximated as $\rho\propto r^{-2}$.

\subsection{Comparison to previous inferences on Tycho}

\sm{\cite{Warren+2013MNRAS} make detailed measurements of the dynamics of their numerical SNR model, also based on an exponential density profile for the ejecta. Upon comparison of the model to ejecta velocities measured from Fe and Si absorption lines in Tycho's remnant, they estimate the scaled age (also called the dynamical age in their work) of Tycho to be in the range $1.0-1.1$. We get a value of $t/T' \approx 1.16$ for Tycho (using Equation \ref{eq:T_WB}, since $T'$ is slightly different from $T$) solely from measurement of the dominant substructures in Tycho, which is close to the estimate obtained from its dynamics. \cite{Katsuda+2010ApJ} use x-ray proper-motion measurements of the forward shock in Tycho in conjunction with the model of \cite{Dwarkadas2000ApJ} to estimate a scaled age of $t/T' \approx 1$. Using this, they deduce a number density of $n_0\approx0.2\;\mathrm{cm^{-3}}$ for Tycho's CSM. Optical \citep{Kirshner+1987ApJ} and x-ray \citep{Cassam-Chenai+2007ApJ} emissivity measurements of Tycho lead to similar estimates ($n_0\sim0.2-0.3\;\mathrm{cm^{-3}}$), consistent with our finding. On the other hand, dust emission modeling using IR flux ratios in Tycho \citep{Williams+2013ApJ} indicates a somewhat lower density ($n_0\sim0.1\;\mathrm{cm^{-3}}$). Nonetheless, most of the observations (of both proper motion and emissivities in Tycho) agree on the presence of relatively denser clumps in the east and consequently an ambient density gradient in the northeast-southwest direction. It is unclear if the low-$l$ modes seen in Tycho's power spectrum probe this gradient; more studies are required to examine such effects.}



\section{Discussion}   \label{sec:discussion}

Our discovery of the typical size of the fleece-like structures in Tycho's remnant is quite robust, although some findings remain unresolved. Firstly, the clump size in Tycho is too big to be consistent with even a modestly lowered adiabatic index of $\gamma=3/2$ (or $4/3$), although an effective adiabatic index somewhere between $5/3$ and $3/2$ cannot be ruled out. This seems to be in some tension with the accepted explanation for Rayleigh-Taylor fingers in Tycho coming in contact with the forward shock. The forward shock is thought to be slowing down due to cooling from particle acceleration, which effectively lowers the adiabatic index of the fluid \citep{Blondin+2001ApJ}. A systematic study with an explicit prescription to describe particle acceleration at the shock \citep[e.g.,][]{Ferrand+2010A&A} is required to investigate this tension further. Another feature of Tycho's power spectrum is its steep slope (approximately $C_l \propto l^{-3}$) at small scales, which is explained well by our $\gamma=5/3$ model. Unfortunately, this shape is also consistent with and thus does not rule out the $\gamma=4/3$ model (Figure \ref{fig:cooling_spectra}). There does exist another major indicator of the effective adiabatic index in our models, namely, the onset of a Kolmogorov turbulent cascade at small scales. At higher particle acceleration efficiency or lower adiabatic indices, the angular power spectrum of density starts showing signature of a turbulent cascade ($C_l \propto l^{-5/3}$) at an earlier age (compare density power spectra in Figures \ref{fig:fiducial_spectra} and \ref{fig:cooling_spectra}). But no such signature is found in the power spectrum of either the synthetic image (Figure \ref{fig:im_spectra_all_models}) or of Tycho's x-ray image. We can therefore hope to capture Kolmogorov turbulence in older remnants only by power spectral analysis of their 3D reconstructed structure. This is a prime example of how images irrecoverably lose information about the remnant and therefore cannot eliminate the necessity of measuring power spectra from 3D reconstructions of remnants.

Secondly, our models are inadequate in explaining the large scale asymmetries in a Type-Ia SNR like Tycho. Additional physics has been invoked to explain these asymmetries, particularly for Tycho, where large scale asymmetries like the bright arc in the northwestern part can be spotted easily. \cite{Vigh+2011ApJ} explore several different possibilities, such as the presence of a large scale density gradient in the CSM, a kick velocity imparted on the ejecta due to breakup of the binary system, and mass loading of the shock by a companion star. \cite{Ferrand+2022ApJ} study the remnant of an inherently asymmetric explosion via a sub-Chandrasekhar double detonation \citep{Fink+2010A&A} of a white dwarf, also taking into account the effect of the companion on the ejecta. The power spectra of these models have large power at low-$l$ modes, corresponding to large scale asymmetries endemic to the explosion itself. Thus different explosion mechanisms and environments leave their signatures at low-$l$ modes in the power spectrum, which should be identified using more realistic models of SNRs. Comparison of these signatures to the low-$l$ modes of power spectra of remnants will be useful in constraining the explosion mechanism and/or asymmetries in the environment.


Lastly, the proxy x-ray images from our models are helpful for demonstrating that they capture the dominant angular mode and small scale behavior of the remnant correctly, but they may not be accurate representations of actual images of remnants. Our assumption of the emissivity of a shocked plasma element being proportional only to the density squared overestimates the emission in general. It holds only when the electrons and ions in the shocked plasma are in temperature equilibrium with each other, which requires sufficiently high densities and enough time to have elapsed since the plasma got shocked. A more accurate image may be produced by explicitly calculating emissivity under conditions of Non-equilibrium Ionization \citep[NEI; see e.g.,][]{Dwarkadas+2010MNRAS,Orlando+2015ApJ}. Studies involving more realistic models and more accurate calculations of thermal x-ray emissivities will be pursued in the future to compare against high resolution images of galactic SNRs.

\section{Conclusion}
\label{sec:conclusion}

\sm{In this work, we demonstrate a way to characterize turbulent activity in SNRs by calculating a power spectrum from their image and connecting them to numerical models. Our models, starting from a radially symmetric ejecta with an exponential density profile, exhibit a dominant scale of substructures that's dependent on their dynamical age, as found previously \citep{Warren+2013MNRAS}. This dominant scale can be found from the power spectrum of the remnant and is a robust indicator of the density profile of the outermost ejecta. The efficiency of particle acceleration in shocks also affects this scale, making it a potential candidate for probing cosmic ray production in SNRs. Besides the size of dominant substructures, the power spectrum also reveals the nature of small scale turbulence and large scale asymmteries, the latter being likely indicators of asymmetric explosion mechanisms and/or progenitor surroundings \citep{Polin+2022,Mandal+2023ApJ}. We apply the technique to Tycho's remnant and successfully obtain the typical size of its fleece-like structures. Our key novel findings are listed below.}


\begin{enumerate}
    \item The power spectra of proxy x-ray images (using $\Delta$-variance) correctly picks out the true dominant angular mode of the remnants. Although we don't have a clear physical interpretation of the shape of the proxy image power spectrum, but it matches the power spectrum of Tycho's remnant at large-$l$ modes very accurately.

    \item The fleece-like structures in Tycho's remnant correspond to an angular harmonic $l_0\approx23$, and implies a scaled age of $t/T\approx0.5$ for Tycho. This is used to infer that Tycho's SNR is expanding against stellar ejecta with a shallow slope near the reverse shock ($\rho(r)\propto r^{-2}$). Assuming ejecta mass and the energy of explosion, the value of $l_0$ implies a number density $\sim0.28\mathrm{\;cm^{-3}}$ for the ambient medium.

    \item Emissivity distribution of Tycho has significant contribution from small angular harmonics (or large scale asymmetries), despite having an overall spherical shape. These are likely endemic to the SN explosion, but may also be caused by external factors like a gradient in the CSM density.

    \item The angular power spectra of density of our SNR models have a steep shape for $l>l_0$ at early times. Later the density power spectra transition to the Kolmogorov power-law shape $C_l \propto l^{-5/3}$ (for $l>l_0$), indicating the presence of a turbulent cascade. \sm{We therefore predict the Kolmogorov power-law shape to be an identifying feature of dynamically older remnants. However, this shape may only be revealed by extracting power spectra from 3D reconstructed maps of the remnant and not from its image.}

    \item The exact time of transition to Kolmogorov turbulence depends on efficiency of particle acceleration in the shock. For more efficient acceleration (lower effective adiabatic index), the angular power spectrum of density transitions to Kolmogorov turbulence earlier.

    \item The efficiency of particle acceleration also affects $l_0$. More efficient acceleration seems to lead to a smaller value of $l_0$ (larger RTI structures) at the same dynamical age.
\end{enumerate}


\sm{The break harmonic ($l_0$) of the power spectrum is a robust indicator of the remnant's scaled (or dynamical) age, even in the presence of asymmetries in the progenitor system. This can be used to infer the density structure of the outermost ejecta, and also an independent measure of the CSM density, provided ejecta mass and explosion energy of the supernova are known. Moreover, the break harmonic may be measured from images of different bands to study substructure sizes for different ions, as done previously by \cite{Lopez+2011ApJ}. Relative sizes of substructures for different elements should be indicative of mixing in the ejecta, as well as the explosion mechanism. Additionally, as seen for Tycho, there may be additional power at small harmonic modes in the power spectrum which may only be explained by and bear signatures of asymmetries endemic to the SN explosion as well as in the surrounding medium. More systematic studies exploring different explosion mechanisms \citep[e.g.,][]{Ferrand+2019ApJ,Ferrand+2021ApJ} and progenitor environments \citep[e.g.,][]{Williams+2013ApJ, Orlando+2022A&A} are needed to identify such signatures accurately and tie them back to the progenitor system.}

\acknowledgments

Numerical calculations were performed on the Petunia computing cluster hosted by the Department of Physics and Astronomy at Purdue University.

\software{\sprout\, \citep{Mandal+2023_sprout},  
    VisIt \citep{HPV:VisIt}, 
    SHTOOLS \citep{SHTOOLS},
    NumPy \citep{numpy},
    Matplotlib \citep{matplotlib}.
}




\label{fig:convergence_full}

\bibliographystyle{apj} 
\typeout{}
\bibliography{smbib}

\begin{thebibliography}{}
\expandafter\ifx\csname natexlab\endcsname\relax\def\natexlab#1{#1}\fi

\bibitem[{{Ar{\'e}valo} {et~al.}(2012){Ar{\'e}valo}, {Churazov}, {Zhuravleva},
  {Hern{\'a}ndez-Monteagudo}, \& {Revnivtsev}}]{Arevalo+2012MNRAS}
{Ar{\'e}valo}, P., {Churazov}, E., {Zhuravleva}, I.,
  {Hern{\'a}ndez-Monteagudo}, C., \& {Revnivtsev}, M. 2012, \mnras, 426, 1793

\bibitem[{{Bear} {et~al.}(2017){Bear}, {Grichener}, \&
  {Soker}}]{Bear+2017MNRAS}
{Bear}, E., {Grichener}, A., \& {Soker}, N. 2017, \mnras, 472, 1770

\bibitem[{Blondin(2005)}]{Blondin2005}
Blondin, J.~M. 2005, Journal of Physics: Conference Series, 16, 370

\bibitem[{Blondin \& Ellison(2001)}]{Blondin+2001ApJ}
Blondin, J.~M., \& Ellison, D.~C. 2001, The Astrophysical Journal, 560, 244

\bibitem[{{Blondin} \& {Mezzacappa}(2007)}]{Blondin+2007Nature}
{Blondin}, J.~M., \& {Mezzacappa}, A. 2007, \nat, 445, 58

\bibitem[{{Cassam-Chena{\"\i}} {et~al.}(2007){Cassam-Chena{\"\i}}, {Hughes},
  {Ballet}, \& {Decourchelle}}]{Cassam-Chenai+2007ApJ}
{Cassam-Chena{\"\i}}, G., {Hughes}, J.~P., {Ballet}, J., \& {Decourchelle}, A.
  2007, \apj, 665, 315

\bibitem[{Celli {et~al.}(2019)Celli, Morlino, Gabici, \&
  Aharonian}]{Celli+2019MNRAS}
Celli, S., Morlino, G., Gabici, S., \& Aharonian, F.~A. 2019, Monthly Notices
  of the Royal Astronomical Society, 487, 3199

\bibitem[{{Charlebois} {et~al.}(2010){Charlebois}, {Drissen}, {Bernier},
  {Grandmont}, \& {Binette}}]{Charlebois+2010AJ}
{Charlebois}, M., {Drissen}, L., {Bernier}, A.~P., {Grandmont}, F., \&
  {Binette}, L. 2010, \aj, 139, 2083

\bibitem[{{Chevalier} {et~al.}(1992){Chevalier}, {Blondin}, \&
  {Emmering}}]{Chevalier+1992ApJ}
{Chevalier}, R.~A., {Blondin}, J.~M., \& {Emmering}, R.~T. 1992, \apj, 392, 118

\bibitem[{{Chevalier} \& {Klein}(1978)}]{Chevalier+1978ApJ}
{Chevalier}, R.~A., \& {Klein}, R.~I. 1978, \apj, 219, 994

\bibitem[{Childs {et~al.}(2012)Childs, Brugger, Whitlock, Meredith, Ahern,
  Pugmire, Biagas, Miller, Harrison, Weber, Krishnan, Fogal, Sanderson, Garth,
  Bethel, Camp, R\"{u}bel, Durant, Favre, \& Navr\'{a}til}]{HPV:VisIt}
Childs, H., Brugger, E., Whitlock, B., {et~al.} 2012, in High Performance
  Visualization--Enabling Extreme-Scale Scientific Insight (Chapman and
  Hall/CRC), 357--372

\bibitem[{{Churazov} {et~al.}(2012){Churazov}, {Vikhlinin}, {Zhuravleva},
  {Schekochihin}, {Parrish}, {Sunyaev}, {Forman}, {B{\"o}hringer}, \&
  {Randall}}]{Churazov+2012MNRAS}
{Churazov}, E., {Vikhlinin}, A., {Zhuravleva}, I., {et~al.} 2012, \mnras, 421,
  1123

\bibitem[{{DeLaney} {et~al.}(2010){DeLaney}, {Rudnick}, {Stage}, {Smith},
  {Isensee}, {Rho}, {Allen}, {Gomez}, {Kozasa}, {Reach}, {Davis}, \&
  {Houck}}]{DeLaney+2010ApJ}
{DeLaney}, T., {Rudnick}, L., {Stage}, M.~D., {et~al.} 2010, \apj, 725, 2038

\bibitem[{{Duffell} \& {MacFadyen}(2011)}]{Duffell+2011ApJS}
{Duffell}, P.~C., \& {MacFadyen}, A.~I. 2011, \apjs, 197, 15

\bibitem[{{Dwarkadas}(2000)}]{Dwarkadas2000ApJ}
{Dwarkadas}, V.~V. 2000, \apj, 541, 418

\bibitem[{{Dwarkadas} \& {Chevalier}(1998)}]{Dwarkadas+1998ApJ}
{Dwarkadas}, V.~V., \& {Chevalier}, R.~A. 1998, \apj, 497, 807

\bibitem[{{Dwarkadas} {et~al.}(2010){Dwarkadas}, {Dewey}, \&
  {Bauer}}]{Dwarkadas+2010MNRAS}
{Dwarkadas}, V.~V., {Dewey}, D., \& {Bauer}, F. 2010, \mnras, 407, 812

\bibitem[{{Ferrand} {et~al.}(2010){Ferrand}, {Decourchelle}, {Ballet},
  {Teyssier}, \& {Fraschetti}}]{Ferrand+2010A&A}
{Ferrand}, G., {Decourchelle}, A., {Ballet}, J., {Teyssier}, R., \&
  {Fraschetti}, F. 2010, \aap, 509, L10

\bibitem[{{Ferrand} {et~al.}(2022){Ferrand}, {Tanikawa}, {Warren}, {Nagataki},
  {Safi-Harb}, \& {Decourchelle}}]{Ferrand+2022ApJ}
{Ferrand}, G., {Tanikawa}, A., {Warren}, D.~C., {et~al.} 2022, \apj, 930, 92

\bibitem[{{Ferrand} {et~al.}(2019){Ferrand}, {Warren}, {Ono}, {Nagataki},
  {R{\"o}pke}, \& {Seitenzahl}}]{Ferrand+2019ApJ}
{Ferrand}, G., {Warren}, D.~C., {Ono}, M., {et~al.} 2019, \apj, 877, 136

\bibitem[{{Ferrand} {et~al.}(2021){Ferrand}, {Warren}, {Ono}, {Nagataki},
  {R{\"o}pke}, {Seitenzahl}, {Lach}, {Iwasaki}, \& {Sato}}]{Ferrand+2021ApJ}
---. 2021, \apj, 906, 93

\bibitem[{{Fesen} \& {Milisavljevic}(2016)}]{Fesen+2016ApJ}
{Fesen}, R.~A., \& {Milisavljevic}, D. 2016, \apj, 818, 17

\bibitem[{{Fink} {et~al.}(2010){Fink}, {R{\"o}pke}, {Hillebrandt},
  {Seitenzahl}, {Sim}, \& {Kromer}}]{Fink+2010A&A}
{Fink}, M., {R{\"o}pke}, F.~K., {Hillebrandt}, W., {et~al.} 2010, \aap, 514,
  A53

\bibitem[{{Gabler} {et~al.}(2021){Gabler}, {Wongwathanarat}, \&
  {Janka}}]{Gabler+2021MNRAS}
{Gabler}, M., {Wongwathanarat}, A., \& {Janka}, H.-T. 2021, \mnras, 502, 3264

\bibitem[{{Gonz{\'a}lez-Casanova} {et~al.}(2014){Gonz{\'a}lez-Casanova}, {De
  Colle}, {Ramirez-Ruiz}, \& {Lopez}}]{GC+2014ApJ}
{Gonz{\'a}lez-Casanova}, D.~F., {De Colle}, F., {Ramirez-Ruiz}, E., \& {Lopez},
  L.~A. 2014, \apjl, 781, L26

\bibitem[{Harris {et~al.}(2020)Harris, Millman, van~der Walt, Gommers,
  Virtanen, Cournapeau, Wieser, Taylor, Berg, Smith, Kern, Picus, Hoyer, van
  Kerkwijk, Brett, Haldane, del R{\'{i}}o, Wiebe, Peterson,
  G{\'{e}}rard-Marchant, Sheppard, Reddy, Weckesser, Abbasi, Gohlke, \&
  Oliphant}]{numpy}
Harris, C.~R., Millman, K.~J., van~der Walt, S.~J., {et~al.} 2020, Nature, 585,
  357

\bibitem[{Hunter(2007)}]{matplotlib}
Hunter, J.~D. 2007, Computing in Science \& Engineering, 9, 90

\bibitem[{{Iwakami} {et~al.}(2008){Iwakami}, {Kotake}, {Ohnishi}, {Yamada}, \&
  {Sawada}}]{Iwakami+2008ApJ}
{Iwakami}, W., {Kotake}, K., {Ohnishi}, N., {Yamada}, S., \& {Sawada}, K. 2008,
  \apj, 678, 1207

\bibitem[{{Katsuda} {et~al.}(2010){Katsuda}, {Petre}, {Hughes}, {Hwang},
  {Yamaguchi}, {Hayato}, {Mori}, \& {Tsunemi}}]{Katsuda+2010ApJ}
{Katsuda}, S., {Petre}, R., {Hughes}, J.~P., {et~al.} 2010, \apj, 709, 1387

\bibitem[{{Kirshner} {et~al.}(1987){Kirshner}, {Winkler}, \&
  {Chevalier}}]{Kirshner+1987ApJ}
{Kirshner}, R., {Winkler}, P.~F., \& {Chevalier}, R.~A. 1987, \apjl, 315, L135

\bibitem[{{Kj{\ae}r} {et~al.}(2010){Kj{\ae}r}, {Leibundgut}, {Fransson},
  {Jerkstrand}, \& {Spyromilio}}]{Kjaer+2010A&A}
{Kj{\ae}r}, K., {Leibundgut}, B., {Fransson}, C., {Jerkstrand}, A., \&
  {Spyromilio}, J. 2010, \aap, 517, A51

\bibitem[{{Kolmogorov}(1941)}]{Kolmogorov1941DoSSR}
{Kolmogorov}, A. 1941, Akademiia Nauk SSSR Doklady, 30, 301

\bibitem[{{Larsson} {et~al.}(2021){Larsson}, {Sollerman}, {Lyman},
  {Spyromilio}, {Tenhu}, {Fransson}, \& {Lundqvist}}]{Larsson+2021ApJ}
{Larsson}, J., {Sollerman}, J., {Lyman}, J.~D., {et~al.} 2021, \apj, 922, 265

\bibitem[{{Law} {et~al.}(2020){Law}, {Milisavljevic}, {Patnaude}, {Plucinsky},
  {Gladders}, {Schmidt}, {Sravan}, {Banovetz}, {Sano}, {McGraw}, {Takahashi},
  \& {Orlando}}]{Law+2020ApJ}
{Law}, C.~J., {Milisavljevic}, D., {Patnaude}, D.~J., {et~al.} 2020, \apj, 894,
  73

\bibitem[{{Lopez} {et~al.}(2013){Lopez}, {Ramirez-Ruiz}, {Castro}, \&
  {Pearson}}]{Lopez+2013ApJ}
{Lopez}, L.~A., {Ramirez-Ruiz}, E., {Castro}, D., \& {Pearson}, S. 2013, \apj,
  764, 50

\bibitem[{{Lopez} {et~al.}(2011){Lopez}, {Ramirez-Ruiz}, {Huppenkothen},
  {Badenes}, \& {Pooley}}]{Lopez+2011ApJ}
{Lopez}, L.~A., {Ramirez-Ruiz}, E., {Huppenkothen}, D., {Badenes}, C., \&
  {Pooley}, D.~A. 2011, \apj, 732, 114

\bibitem[{{Lopez} {et~al.}(2009){Lopez}, {Ramirez-Ruiz}, {Pooley}, \&
  {Jeltema}}]{Lopez+2009ApJ}
{Lopez}, L.~A., {Ramirez-Ruiz}, E., {Pooley}, D.~A., \& {Jeltema}, T.~E. 2009,
  \apj, 691, 875

\bibitem[{Mandal \& Duffell(2023)}]{Mandal+2023_sprout}
Mandal, S., \& Duffell, P.~C. 2023, The Astrophysical Journal Supplement
  Series, 269, 30

\bibitem[{{Mandal} {et~al.}(2023){Mandal}, {Duffell}, {Polin}, \&
  {Milisavljevic}}]{Mandal+2023ApJ}
{Mandal}, S., {Duffell}, P.~C., {Polin}, A., \& {Milisavljevic}, D. 2023, \apj,
  956, 130

\bibitem[{{Martin} {et~al.}(2021){Martin}, {Milisavljevic}, \&
  {Drissen}}]{Martin+2021MNRAS}
{Martin}, T., {Milisavljevic}, D., \& {Drissen}, L. 2021, \mnras, 502, 1864

\bibitem[{{McCray} \& {Fransson}(2016)}]{McCray+2016ARA&A}
{McCray}, R., \& {Fransson}, C. 2016, \araa, 54, 19

\bibitem[{{Milisavljevic} \& {Fesen}(2013)}]{Milisavljevic+2013}
{Milisavljevic}, D., \& {Fesen}, R.~A. 2013, \apj, 772, 134

\bibitem[{{Milisavljevic} \& {Fesen}(2015)}]{Milisavljevic+2015}
---. 2015, Science, 347, 526

\bibitem[{{Millard} {et~al.}(2022){Millard}, {Park}, {Sato}, {Hughes}, {Slane},
  {Patnaude}, {Burrows}, \& {Badenes}}]{Millard+2022ApJ}
{Millard}, M.~J., {Park}, S., {Sato}, T., {et~al.} 2022, \apj, 937, 121

\bibitem[{{Nomoto} {et~al.}(1984){Nomoto}, {Thielemann}, \&
  {Yokoi}}]{Nomoto+1984ApJ}
{Nomoto}, K., {Thielemann}, F.~K., \& {Yokoi}, K. 1984, \apj, 286, 644

\bibitem[{{Orlando} {et~al.}(2015){Orlando}, {Miceli}, {Pumo}, \&
  {Bocchino}}]{Orlando+2015ApJ}
{Orlando}, S., {Miceli}, M., {Pumo}, M.~L., \& {Bocchino}, F. 2015, \apj, 810,
  168

\bibitem[{{Orlando} {et~al.}(2016){Orlando}, {Miceli}, {Pumo}, \&
  {Bocchino}}]{Orlando+2016ApJ}
---. 2016, \apj, 822, 22

\bibitem[{{Orlando} {et~al.}(2021){Orlando}, {Wongwathanarat}, {Janka},
  {Miceli}, {Ono}, {Nagataki}, {Bocchino}, \& {Peres}}]{Orlando+2021A&A}
{Orlando}, S., {Wongwathanarat}, A., {Janka}, H.~T., {et~al.} 2021, \aap, 645,
  A66

\bibitem[{{Orlando} {et~al.}(2022){Orlando}, {Wongwathanarat}, {Janka},
  {Miceli}, {Nagataki}, {Ono}, {Bocchino}, {Vink}, {Milisavljevic}, {Patnaude},
  \& {Peres}}]{Orlando+2022A&A}
---. 2022, \aap, 666, A2

\bibitem[{{Ossenkopf} {et~al.}(2008){Ossenkopf}, {Krips}, \&
  {Stutzki}}]{Ossenkopf+2008A&A}
{Ossenkopf}, V., {Krips}, M., \& {Stutzki}, J. 2008, \aap, 485, 917

\bibitem[{Polin {et~al.}(2022)Polin, Duffell, \& Milisavljevic}]{Polin+2022}
Polin, A., Duffell, P., \& Milisavljevic, D. 2022, The Astrophysical Journal
  Letters, 940, L28

\bibitem[{{Reynolds} {et~al.}(2007){Reynolds}, {Borkowski}, {Hwang}, {Hughes},
  {Badenes}, {Laming}, \& {Blondin}}]{Reynolds+2007ApJ}
{Reynolds}, S.~P., {Borkowski}, K.~J., {Hwang}, U., {et~al.} 2007, \apjl, 668,
  L135

\bibitem[{{Sano} {et~al.}(2020{\natexlab{a}}){Sano}, {Plucinsky}, {Bamba},
  {Sharda}, {Filipovi{\'c}}, {Law}, {Alsaberi}, {Yamane}, {Tokuda}, {Acero},
  {Sasaki}, {Vink}, {Inoue}, {Inutsuka}, {Shimoda}, {Tsuge}, {Fujii}, {Voisin},
  {Maxted}, {Rowell}, {Onishi}, {Kawamura}, {Mizuno}, {Yamamoto}, {Tachihara},
  \& {Fukui}}]{Sano+2020ApJ_a}
{Sano}, H., {Plucinsky}, P.~P., {Bamba}, A., {et~al.} 2020{\natexlab{a}}, \apj,
  902, 53

\bibitem[{{Sano} {et~al.}(2020{\natexlab{b}}){Sano}, {Inoue}, {Tokuda},
  {Tanaka}, {Yamazaki}, {Inutsuka}, {Aharonian}, {Rowell}, {Filipovi{\'c}},
  {Yamane}, {Yoshiike}, {Maxted}, {Uchida}, {Hayakawa}, {Tachihara},
  {Uchiyama}, \& {Fukui}}]{Sano+2020ApJ_b}
{Sano}, H., {Inoue}, T., {Tokuda}, K., {et~al.} 2020{\natexlab{b}}, \apjl, 904,
  L24

\bibitem[{{Springel}(2010)}]{Springel2010MNRAS}
{Springel}, V. 2010, \mnras, 401, 791

\bibitem[{{Uchida} {et~al.}(2024){Uchida}, {Kasuga}, {Maeda}, {Lee}, {Tanaka},
  \& {Bamba}}]{Uchida+2024arXiv}
{Uchida}, H., {Kasuga}, T., {Maeda}, K., {et~al.} 2024, arXiv e-prints,
  arXiv:2401.11763

\bibitem[{{Velazquez} {et~al.}(1998){Velazquez}, {Gomez}, {Dubner}, {de
  Castro}, \& {Costa}}]{Velazquez+1998A&A}
{Velazquez}, P.~F., {Gomez}, D.~O., {Dubner}, G.~M., {de Castro}, G.~G., \&
  {Costa}, A. 1998, \aap, 334, 1060

\bibitem[{{Vigh} {et~al.}(2011){Vigh}, {Vel{\'a}zquez}, {G{\'o}mez}, {Reynoso},
  {Esquivel}, \& {Matias Schneiter}}]{Vigh+2011ApJ}
{Vigh}, C.~D., {Vel{\'a}zquez}, P.~F., {G{\'o}mez}, D.~O., {et~al.} 2011, \apj,
  727, 32

\bibitem[{{Vogt} \& {Dopita}(2010)}]{Vogt+2010ApJ}
{Vogt}, F., \& {Dopita}, M.~A. 2010, \apj, 721, 597

\bibitem[{{Vogt} \& {Dopita}(2011)}]{Vogt+2011ApJ}
---. 2011, \apss, 331, 521

\bibitem[{{Warren} \& {Blondin}(2013)}]{Warren+2013MNRAS}
{Warren}, D.~C., \& {Blondin}, J.~M. 2013, \mnras, 429, 3099

\bibitem[{{Warren} {et~al.}(2005){Warren}, {Hughes}, {Badenes}, {Ghavamian},
  {McKee}, {Moffett}, {Plucinsky}, {Rakowski}, {Reynoso}, \&
  {Slane}}]{Warren+2005ApJ}
{Warren}, J.~S., {Hughes}, J.~P., {Badenes}, C., {et~al.} 2005, \apj, 634, 376

\bibitem[{Wieczorek \& Meschede(2018)}]{SHTOOLS}
Wieczorek, M.~A., \& Meschede, M. 2018, Geochemistry, Geophysics, Geosystems,
  19, 2574

\bibitem[{{Williams} {et~al.}(2013){Williams}, {Borkowski}, {Ghavamian},
  {Hewitt}, {Mao}, {Petre}, {Reynolds}, \& {Blondin}}]{Williams+2013ApJ}
{Williams}, B.~J., {Borkowski}, K.~J., {Ghavamian}, P., {et~al.} 2013, \apj,
  770, 129

\bibitem[{{Winkler} {et~al.}(2003){Winkler}, {Gupta}, \&
  {Long}}]{Winkler+2003ApJ}
{Winkler}, P.~F., {Gupta}, G., \& {Long}, K.~S. 2003, \apj, 585, 324

\bibitem[{{Wongwathanarat} {et~al.}(2017){Wongwathanarat}, {Janka},
  {M{\"u}ller}, {Pllumbi}, \& {Wanajo}}]{Wongwathanarat+2017ApJ}
{Wongwathanarat}, A., {Janka}, H.-T., {M{\"u}ller}, E., {Pllumbi}, E., \&
  {Wanajo}, S. 2017, \apj, 842, 13

\bibitem[{{Wongwathanarat} {et~al.}(2015){Wongwathanarat}, {M{\"u}ller}, \&
  {Janka}}]{Wongwathanarat+2015A&A}
{Wongwathanarat}, A., {M{\"u}ller}, E., \& {Janka}, H.~T. 2015, \aap, 577, A48

\end{thebibliography}

\end{document}